\documentclass[english]{iopart}

\usepackage{graphicx}
\usepackage{url}
\usepackage{iopams} 

\expandafter\let\csname equation*\endcsname\relax
\expandafter\let\csname endequation*\endcsname\relax

\usepackage{amssymb}
\usepackage[latin1]{inputenc}
\usepackage{amsmath}
\usepackage{epsfig}
\newcounter{fig}
\begin{document}

\title{The Ising correlation $C(M,N)$ for $\nu=-k$}
      
\vskip .1cm 

\author{S. Boukraa$^\dag$,  
J.-M. Maillard$^\ddag$, B. M. McCoy$^\P$,}
\address{\dag IAESB,
 Universit\'e de Blida, Algeria}
\address{\ddag\ LPTMC, Sorbonne Universit\'e,  Tour 24, 5\`eme \'etage, case 121, \\
 4 Place Jussieu, 75252 Paris Cedex 05, France} 
\address{\P Institute for Theoretical Physics,
State University of New York,
Stony Brook, USA}
\ead{maillard@lptmc.jussieu.fr, jean-marie.maillard@sorbonne-universite.fr, 
bkrsalah@yahoo.com, mccoy@max2.physics.sunysb.edu}

\vskip .1cm 

\begin{abstract}

We present  Painlev{\'e} VI sigma form equations for the general Ising
low and high temperature two-point correlation functions $ \, C(M,N)$  with
$\, M \, \leq \, N \, $ in the special case $\nu = \, -k$
where $\nu = \, \sinh 2E_h/k_BT/\sinh 2E_v/k_BT$. More specifically
four different non-linear ODEs depending explicitly
on the two integers $\,M $ and $\, N$ emerge: these four non-linear ODEs 
correspond  to distinguish respectively low and high temperature, 
together with  $\, M+N$ even or odd. These four different non-linear ODEs
are also valid for $\, M \, \ge\, N$ when $\, \nu \, = \, -1/k$.
For the low-temperature row correlation functions $ \, C(0,N)$ with $\, N$ odd, 
we exhibit again for this selected  $\nu = \, -k$ condition, a remarkable
phenomenon of a Painlev\'e VI sigma function being the sum of four Painlev\'e VI
sigma functions having the same Okamoto parameters.
We show in this $\nu = \, -k$ case
 for $\, T <\, T_c \, $ and also $\, T > \, T_c$, 
that $\, C(M,N)$ with $\, M \, \leq \, N \, $ 
is given as an $ \, N \times\,  N$ Toeplitz determinant.  

\end{abstract}

\vskip .1cm

\noindent {\bf PACS}: 05.50.+q, 05.10.-a, 02.30.Hq, 02.30.Gp, 02.40.Xx

\noindent {\bf AMS Classification scheme numbers}: 34M55, 
47E05, 81Qxx, 32G34, 34Lxx, 34Mxx, 14Kxx 

\vskip .2cm

{\bf Key-words}: Ising correlation functions, sigma form of Painlev\'e VI,
Okamoto parameters, Toeplitz determinants.

\vskip .2cm
\vskip .2cm

\section{Introduction}
\label{Introduction}

The anisotropic Ising model on the square lattice is defined
by the interaction energy
\begin{equation}
  \hspace{-0.7in} \quad 
         {\mathcal E \,}
\, \, = \, \,\, -\sum_{j,k}\{E_v\sigma_{j,k}\sigma_{j+1,k}
\, +E_h\sigma_{j,k}\sigma_{j,k+1}\}, 
\end{equation}
where $\sigma_{j,k}\, = \,\, \pm 1$ is the spin at row $\, j$
and column $\, k$ and the sum is over all lattice sites. The free
energy in the thermodynamic limit was computed by Onsager~\cite{ons} in 1944. 

The investigation of the 
correlation functions was initiated by Kaufman and Onsager~\cite{kauf}
in 1949 and in 1963 Montroll, Potts and Ward~\cite{mpw}
extended and simplified these results to show that all correlations can
be written as determinants in an infinite number of ways. The simplest
of these cases is the row correlation
\begin{equation}
\hspace{-0.3in}
C(0,N)\, = \,\,  \langle \sigma_{0,0} \sigma_{0,N}\rangle
\,  \, = \, \,  \, \, 
\begin{array}{|llll|}
{a}_0&{a}_{-1}& \cdots &{a}_{-N+1}\\
{a}_1&{ a}_0& \cdots &{a}_{-N+2}\\
\vdots & \vdots && \vdots\\
{a}_{N-1}& {a}_{N-2}& \cdots &{a}_0
\label{dn}
\end{array} \, , 
\end{equation}
with
\begin{equation}
\label{an}
\hspace{-0.6in} \quad 
a_n\,\, = \,\,\, \,  {1\over 2\pi} \, \, \int_{0}^{2\pi} \,
\left[ {(1\,-\alpha_1 \, e^{i\theta})\, (1\, -\alpha_2 \, e^{-i\theta})\over
(1 \, -\alpha_1 \, e^{-i\theta}) \, (1 \, -\alpha_2 \,e^{i\theta})}\right]^{1/2}
  \, e^{in\theta} \, \,  d\theta, 
\end{equation}
where 
\begin{equation}
\hspace{-0.6in} \quad \quad
\alpha_1 \, = \, \,  e^{-2E_v/k_BT} \,\tanh E_h/k_BT, \quad \quad
 \alpha_2\, = \,\,  e^{-2E_v/k_BT} \,\coth E_h/k_BT, 
\end{equation}
and the diagonal correlation $\, C(N,N)$ also given
by (\ref{dn}) and (\ref{an}) with
\begin{equation}
 \hspace{-0.6in}  \quad
\alpha_1 \, =\, \, 0, \quad \quad
\alpha_2 \,=\, \, k \,=\, \, (\sinh 2E_v/k_BT \,\sinh 2E_h/k_BT)^{-1}. 
\end{equation}
Both the free energy and the correlations have singularites at the
critical temperature $\,T_c$ defined by
\begin{equation}
\hspace{-0.6in}
k\,= \,\, (\sinh 2E_v/k_BT_c \, \sinh 2E_h/k_BT_c)^{-1} \, = \,\, 1. 
\end{equation}

In 1976 Wu, McCoy, Tracy and Barouch~\cite{wmtb} discovered,  in
the scaling limit $ \, T \rightarrow \, T_c\,$ with $ \, N \cdot \, (T \, -T_c) \, $
fixed, that the diagonal correlation $\, C(N,N)$ is given
by a Painlev{\'e} III function.
This was generalized in 1980 by Jimbo and Miwa~\cite{jm}
who defined for $\,T< \,T_c$ 
\begin{equation}
\label{sigmam}
\hspace{-0.4in}
\sigma\, \,= \,\,\,
t \cdot \, (t-1) \cdot \, \frac{d}{dt}\ln C(N,N) \, \, -\frac{t}{4}
\quad \, \quad  {\rm with} \quad \quad \, t \, = \,\, k^2, 
\end{equation}
and for $\,T \, > \,T_c$
\begin{equation}
\label{sigmap}
\sigma \,\,= \,\,\,
t \cdot \, (t-1)\cdot  \, \frac{d}{dt}\ln C(N,N)\, \, -\frac{1}{4}
\quad \,\quad   {\rm with} \quad \quad \,  t \, = \, \, k^{-2}, 
\end{equation}
 and in both cases derived:
\begin{eqnarray}
\label{jmequation}
\hspace{-0.98in}&& \quad  
\left(t \cdot \, (t-1) \cdot \, \frac{d^2\sigma}{dt^2}\right)^2
\\
\hspace{-0.98in}&&  \quad  \,    \quad  
\, = \, \,\, N^2 \cdot \,
\left((t -1) \cdot  \, \frac{d\sigma}{dt} \, -\sigma\right)^2
\, -4 \cdot \,\frac{d\sigma}{dt} \cdot \,
\left((t -1) \cdot \frac{d\sigma}{dt}\, -\sigma\, -\frac{1}{4}\right)
\cdot \,\left(t\frac{d\sigma}{dt}-\sigma\right).
\nonumber 
\end{eqnarray}

For $\, T\, < \, T_c\, $ the boundary condition for (\ref{jmequation})
at $\,t =\, 0\, $ is 
\begin{equation}
\label{bcm}
\hspace{-0.01in} \quad 
C(N,N; \, t) \, \, = \, \,\, 
(1-t)^{1/4} \cdot \,
\Bigl[1\, +\lambda^2 \cdot \,\frac{(1/2)_N  \,(3/2)_N}
{4[(N+1)!]^2} \cdot \, t^{N+1}  \cdot \, (1\,\, +O(t))\Bigr], 
\end{equation}
with $ \,\lambda =\, 1$, $\, (a)_n = \, a (a+1)\cdots (a+n-1)\, $
and $ \,(a)_0= \, 1$. 
For $\, T\, > \, T_c$ the boundary condition at $\,t\, =\, 0$ is
\begin{eqnarray}
\label{bcp}
\hspace{-0.98in}&& \, \,  \quad \quad  \quad 
C(N,N; \, t) \,\,  = \, \, \, 
  (1-t)^{1/4} \cdot \, t^{N/2} \cdot \, \Bigl[\frac{(1/2)_N}{N!}{}
  \cdot \, _2F_1\Bigl([\frac{1}{2},\, N \, +\frac{1}{2}], \, [N+1], \, t\Bigr)
\nonumber\\
\hspace{-0.98in} && \quad \quad \quad  \quad \quad  \quad   \quad  
+\lambda^2 \cdot \,
\frac{(1/2)_N \, ((3/2)_N)^2}{ 16 \, \, (N+1)! \,(N+2)! }
   \cdot \,t^{N+2} \cdot \, (1 \,  \, +O(t))\Bigr], 
\end{eqnarray}
with $\, \lambda = \,1$.

We note for both cases of $\, T \, < \, T_c$ and $\, T \, > \, T_c$
that there are  solutions of (\ref{jmequation}) with boundary condition
where $\,\lambda \, \neq \,  1$.
Those solutions do not correspond to the determinants for
$ \, C(N,N)$ but rather for the lambda extended Fredholm determinants
obtained from the form factor expansions~\cite{lm,Holonomy}. We also
remark that for $ \, T\, > \, T_c$ the term in (\ref{bcm}) with
$\, \lambda \, = \, 0\, $ is by itself an exact solution of (\ref{jmequation})
even though it is not a correlation function of the Ising model.

\vspace{.1in}

{\em It is an outstanding open question to generalize} (\ref{jmequation})
{\em to the general two-point correlation functions
  $\,C(M,N)= \,\langle\sigma_{0,0}\sigma_{M,N}\rangle$.}

\vspace{.1in}

In this paper we consider the correlation $ \, C(M,N)\, $ with anisotropy
\begin{equation}
\hspace{-0.98in}
\nu \, = \, \, \frac{\sinh 2E_h/k_BT}{\sinh 2E_v/k_BT}, 
\end{equation}
for the special case 
\begin{equation}
\label{nukres}  
\hspace{-0.98in}
\nu \,= \, \, -k, 
\end{equation}
which corresponds to
\begin{equation}
\hspace{-0.98in} \quad  \quad  \quad  
\sinh2E_h/k_BT \, = \, \, \pm i, \quad  \quad \quad
k \, = \, \,  \mp \, \frac{i}{\sinh2E_v/k_BT}.
\end{equation}
Because $\, k\, \rightarrow\,  0 \,$ as $\, T\, \rightarrow\,  0$
we refer to this case as $\, T\, <\, T_c \, $ for $\nu$ and $k$
real even though $\, E_v$ and $\, E_h$ are complex
(and hence {\em unphysical}).
 
We also consider the special case
\begin{equation}
\label{nukresp}
\hspace{-0.98in}
\nu \, = \,\, -1/k \, \, = \,\, -k_>, 
\end{equation}
where 
\begin{equation}
\hspace{-0.98in} \quad  \quad  \quad  
\sinh2E_v/k_BT\, = \, \, \pm i,  \quad\quad  \quad
k_> \, =\, \, \mp i \, \, \sinh 2E_h/k_BT
\end{equation}
and because $\,  k_> \, \rightarrow\,  0 \, $ corresponds to
$\, T\, \rightarrow \,  \infty$ we refer to this case as $\, T\, >\, T_c$.
In both cases we find that there is indeed a generalization of (\ref{jmequation}).

For concretness we consider $\, M\,  \geq \,  0\, $ and $\,  N \, \geq \,  0$. We
note that the formalism for $\,  M \, \leq\,  N$ and $ \, M \, \geq\,  N$ is
different but, in the general, the symmetry under $ \, M \, \leftrightarrow \, N$
and $\, E_v \, \leftrightarrow \, E_h \, $ yields the relation:
\begin{equation}
\label{cmnsym}
\hspace{-0.98in}
C(M, \, N; \, k,\, \nu)\, \, = \, \, \, C(N,\, M; \, k, \, 1/\nu).
\end{equation}
However the restrictions (\ref{nukres}) and (\ref{nukresp}) are
{\em not preserved} by (\ref{cmnsym}) and we have instead:
\begin{equation}
\label{mnsym}
\hspace{-0.98in}\quad 
C(M, \, N; \, \nu=-k)\, \,=\, \, \, C(N, \, M; \, \nu= -1/k).
\end{equation}
In this paper we consider only $\, M\, \leq\,  N$ with some remarks about
$\, M\, \geq \, N$ at the end of subsection \ref{subsec} 
and in the discussion section \ref{Discussion}.

We recall previous results~\cite{mjm} on $\,C(M,N)$
in section \ref{correlCMN}. In section \ref{nonlin}   
we use the  program {\it guessfunc}  developed by
Jay Pantone~\cite{Pantone} to find, from large series expansion,  
{\em nonlinear differential equations} for $\, C (M,N) \,$ with
$\, M\, \leq\,  N$, both for
$\,T \, < \, T_c $ (see equations (\ref{eqnm}) below) and for $\, T \, > \,T_c$
(see equations (\ref{EqMNhigh}) and (\ref{eqnp}) below). In section \ref{Oka}
we transform these nonlinear differential equations into the canonical form
of Okamoto~\cite{oka} for sigma form of Painlev{\'e} VI.  
In section \ref{Forrester} we compare our equations with the ones obtained
by Forrester and Witte
for determinants~\cite{fw} as given in~\cite{gil} 
and show in (\ref{cmndn}),  for $\,\nu\,=\,-k \, $ and $\, T \,<\,T_c$,
that $\, C(M,N)$  for $\, M \, \le N$ is expressed as an $\, N \,\times \, N$
Toeplitz determinant. \ref{ToeplitzTsupTc} shows when $\, T \,> \,T_c$ that  $\, C(M,N)$ 
can also be expressed as an $\, N \,\times \, N$ Toeplitz determinant. 
In section \ref{Factoriz} we show for $ T\, < \, T_c$,  $\,\nu\,=\,-k \, $ and $ M+N$ odd
that $\, C(M,N)$, $\, M\, \leq\,  N$, factors, and for $\, C(0,N)$ that
these factors {\em also satisfy} an Okamoto sigma form of
Painlev\'e VI equation. We conclude
in section  \ref{Discussion} with
a discussion of several open questions. 
In  \ref{appA} we give
examples of $\, C(M, N)$ with $\,\nu\,= \, -k \, $ 
for both $\, T\,< \, T_c$ and $\,T > \,T_c$.
 In  \ref{boundaryd}  we present the one parameter family
 of boundary conditions for the general Painlev\'e VI sigma Okamoto
 form which are analytic at $\, k= \, 0$.

\vskip .1cm

\section{The correlation $\, C(M,N)$ for $\, \nu=\,-k$}
\label{correlCMN}

In~\cite{mjm} it was shown  for all $\, M, \, N$ that  the correlation
$\, C(M,N)$ with $\, M \, \le \, N$  can be written
 for all $\, M, \, N$  as a
{\em homogeneous polynomial} in the three elliptic integrals
\begin{eqnarray}
\label{Kdef}
\hspace{-0.48in}
&&{\tilde K}(k)\, \, = \,\,\,
\frac{2}{\pi}\,
\int_0^{\pi/2}\frac{d\theta}{(1 \, -k^2\sin^2\theta)^{1/2}}
\,= \,\,\, {}_2F_1\Bigl([\frac{1}{2},\frac{1}{2}], \, [1], \, \, k^2\Bigr),
\\
\label{Edef}
\hspace{-0.48in}
&&{\tilde E}(k) \,\, = \,\,\, \frac{2}{\pi}
\int_0^{\pi/2}\, d\theta \,\,  (1\, -k^2\sin^2\theta)^{1/2}
\,= \,\,\, {}_2F_1\Bigl([\frac{1}{2},-\frac{1}{2}], \, [1], \, \, k^2\Bigr), 
\\
\label{Pidef}
\hspace{-0.48in}
&&{\tilde \Pi}(-k\nu, \, k) \,\, = \,\,\,
\frac{2}{\pi} \,\int_0^{\pi/2} \,\,
     {{ d\theta} \over  {
         (1\,+k  \, \nu \, \sin^2\theta) \, (1 \, -k^2 \, \sin^2\theta)^{1/2} }}, 
\end{eqnarray}
where 
\begin{eqnarray}
\label{kalpha}
\hspace{-0.58in}
&&k \, \,=\,\, \, (\sinh 2E_v/kT\sinh 2E_h/kT)^{-1} \,= \,\, (s_v \, s_h)^{-1}
\,  =\,\, \frac{\alpha_2 -\alpha_1}{1 \, -\alpha_1\alpha_2}, 
\\
\label{nualpha}
\hspace{-0.58in}
&&\nu \,\,=\,\, \, \frac{\sinh 2E_h/kT}{\sinh2E_v/kT} \,= \,\, \frac{s_h}{s_v}
\,=\,\,\frac{4 \, \alpha_1\alpha_2}{(\alpha_2 -\alpha_1)(1 \, -\alpha_1 \alpha_2)},
\quad \quad \, \, \,  k \,\nu \,=\,\,\frac{1}{s_v^2}, 
\end{eqnarray}
which are valid for
\begin{equation}
\label{validfor}
\hspace{-0.68in}
0\, \leq \,\, k \,\, \leq\, 1\, \quad  \quad {\rm and}
\quad \quad -1 \, \leq\, k \,\nu, 
\end{equation}
$\, {}_2F_1([a,b],\, [c], \, z) \, $ being the hypergeometric function. 

\vskip .1cm

\subsection{$C(0,1) \, $ for $\, \nu\,=\,\,-k$}

It was shown in~\cite{ons} and~\cite{kauf}, for $ \,T \,< \, T_c$ where
$ \,\alpha_1 \,< \,\alpha_2 \,< \,1$, that
\begin{equation}
\label{c01}
C(0,1) \, \,= \, \, \, \sqrt{ 1 \, +\nu k} \cdot \,
\Bigl[ \Bigl(1 \, +{{k} \over {\nu}}\Bigr) \cdot \, {\tilde \Pi}(-\nu k, \,k)
  \,  \,  -{{k} \over {\nu}} \cdot  \, {\tilde K}(k) \Bigr],
\end{equation}
which is conveniently rewritten as
\begin{equation}
\label{newc01}
\hspace{-0.20in}
C(0,1) \, \,= \, \, \, \sqrt{1 +\nu k} \cdot \,  \frac{2}{\pi}
\, \int_{0}^{\pi/2} \, d\theta \cdot \,
\frac{(1 \,-k^2 \, \sin^2\theta)^{1/2}}{1 \, +k \nu \, \sin^2\theta}. 
\end{equation}
For $ \,T \,> \,T_c \, $ where $ \,1 \,< \,\alpha_2\,$ and $ \,\, k_> \,= \, \,1/k$
\begin{eqnarray}
&&\hspace{-0.78in}
C(0,1) \,= \, \, \frac{1}{\nu} \cdot \,\sqrt{1+\nu/k_>} \cdot \, 
\Bigl[(1+\nu k_>) \cdot \, \tilde{\Pi}(-\nu k_>,k_>) \, \, -\tilde{K}(k_>)\Bigr]
\nonumber \\
\label{c01p}
&&\hspace{-0.78in}
\quad  = \,  \, \, k_> \cdot \, \sqrt{1+\nu/k_>} \cdot \,
\frac{2}{\pi} \,\int_0^{\pi/2} \, 
\frac{1 \, -\sin^2\theta}{(1 \, +k_>\nu \, \sin^2\theta)
  \, (1 \, -k_>^2 \, \sin^2\theta)^{1/2}}
\cdot \,  d\theta. 
\end{eqnarray}
In general $ \, C(M,N)$ depends on two (complex) variables through the elliptic
integral $ \,  {\tilde \Pi}(-k\nu, \, k)$.  However, as is seen in (\ref{dn}) and
(\ref{an}) the row correlation $ \, C(0,N)$ reduces to $ \, C(N,N)$ when 
$\alpha_1 \, = \, 0$ (which from (\ref{nualpha}) is equivalent to $\nu \, = \, 0$)   
because $ \, {\tilde \Pi}(-k\nu, \, k)$ degenerates to $ \,  {\tilde K}(k)$.  

There are two other special cases where $ \, {\tilde \Pi}(-k\nu, \, k)$
reduces to combinations of $ \, {\tilde K}(k)$ and $ \, {\tilde E}(k)$. 
One case is the {\em isotropic} case
$\nu \, = \, 1$ where 
\begin{equation}
\hspace{-0.58in}  
  {\tilde \Pi}(-k, \, k) \, \, = \, \, \, \,
  \frac{1}{2} \cdot \, {\tilde K}(k) \,  \,  +\frac{1}{2 \cdot \, (1\, +k)}. 
\end{equation}
Many examples of this reduction of $\, C(M,N)$ have been
given by Shrock and Ghosh~\cite{shrock1,shrock2}.

Another case of reduction of $\, {\tilde \Pi}(-k\nu, \, k)$ is 
\begin{equation}
\label{numinusk}
\hspace{-0.78in}  
\nu \,= \, \, -k, 
\end{equation}
where 
\begin{equation}
\label{EllipticPinuk}
\hspace{-0.88in}  
{\tilde \Pi}(k^2, \,k) \, \,  = \, \, \, \frac{{\tilde E}(k)}{1 \,-k^2}, 
\end{equation}
and the $ \,C(M,N)$, $\, M \, \le \, N$ {\em are reduced to homogeneous polynomials
of the two ellptic integrals} $\, {\tilde E}(k)$ and $\, {\tilde K}(k)$.
For example for $ \,T \,< \,T_c$ when $ \,\nu \,= \,-k$ we see
from (\ref{newc01}) that
\begin{equation}
\label{specialc01m}
\hspace{-0.21in} \quad  \,\,\,\,\,\,
C(0,1) \,\, = \, \,\,  \sqrt{1-k^2} \cdot \,\frac{2}{\pi} \, \int_0^{\pi/2} \,  \,
\frac{1}{(1 \, -k^2 \, \sin^2\theta)^{1/2}} \cdot \,d\theta
\,\, \,= \,  \,\,\sqrt{1-k^2} \cdot \, {\tilde  K}(k). 
\end{equation}
For $\, T\,>\,T_c\,$ from (\ref{c01p})
when $\,\nu\,=\,\, -k_>\,\, = \,\, -1/k$
\begin{equation}
\label{zero}
\hspace{-0.38in}
C(0,1)\,\, =\,\,\,0, 
\end{equation}
because of the vanishing of the square root factor. 
If we remove this factor by writing for $\,T\,>\,T_c$
\begin{equation}
\label{newc01t}
C(0,1) \, \, = \,\,  \,  \sqrt{1\,+\nu/k_{>}} \cdot \, {\tilde C(0,1)}, 
\end{equation}
we find, in the special case $\, \nu\, = \, -k_{>}$,  that 
\begin{equation}
\label{c01tp}
\hspace{-0.23in}
{\tilde  C(0,1)}
\,\,   =\,\,\,  \,    \frac{k_>}{2} \cdot \,
{}_2F_1([\frac{3}{2},\frac{1}{2}],\, [2], \, \, k^2_>)
\, \,  =\,\, \,  \,
   {{ {\tilde K}(k_>)\, -{\tilde E}(k_>) } \over {k_>}}. 
\end{equation} 

\vskip .1cm

\subsection{$C(0, N)$ for $\, T\, < \,T_c$ at $\,\,\nu \,= \, \, -k$}
\label{subCON}

More generally for $\,T\,< \,T_c \, $ we find, from (\ref{kalpha})
and (\ref{nualpha}), that if $\, \,\nu\,= \,\,-k$ then
\begin{equation}
\label{alpha2}
\hspace{-0.38in}
 \alpha_2 \,= \,\,-\alpha_1 \,=\,\,\alpha
\quad \quad \,  {\rm and}  \quad \quad k\,=\,\,\frac{2\alpha}{\alpha^2+1}, 
\quad \,\,\, s_h \, =\,\, i, \quad \,\,\, s_v\,=\,\, -{{i} \over {k}}, 
\end{equation}
where, for simplicity, we have defined $\,\alpha \,= \,\,\alpha_2$.
Thus for $\,\nu\,=\,-k\, $ the $\,a_n$ matrix elements  (\ref{an})
reduce to
\begin{equation}
\label{antheta}
\hspace{-0.18in}
a_n\,\, = \,\, \,\frac{1}{2\pi} \, \int_{0}^{2\pi} \, 
\frac{1\, -\alpha^2\, +\alpha\, (e^{i\theta}\, -e^{-i\theta})}
{\{(1\,-\alpha^2e^{2i\theta})\,(1-\alpha^2e^{-2i\theta})\}^{1/2}}
 \cdot \,  e^{in\theta} \cdot \, d\theta, 
\end{equation}
which, by sending $\,\alpha\, \rightarrow\, -\alpha \, $
and $\,\theta \,\rightarrow \,-\theta \, $
has the symmetry:
\begin{equation}
\label{anmn1}
\hspace{-0.38in}
a_{-n}(k)\,\, =\,\, \,a_n(-k). 
\end{equation}
By considering invariance under $\, \theta\,\rightarrow\, \theta\, +\pi$
and setting $\,2\theta\,= \,\, \phi$,
we see that
\begin{eqnarray}
\label{ae}
\hspace{-0.38in}
&&a_{2m} \,  \, = \,  \, \,
\frac{1}{2\pi}\,\int_0^{2\pi} \, \,\frac{1-\alpha^2}
{\{(1\,-\alpha^2 \, e^{i\phi})\,(1\,-\alpha^2 \, e^{-i\phi})\}^{1/2}}
\cdot \,  e^{i|m|\phi} \cdot \,  d\,\phi, 
\end{eqnarray}
and:
\begin{eqnarray}
\label{ao}
\hspace{-0.48in}
&&a_{\pm(2|m|+1)}\, =\, \, \,\frac{\pm}{2\pi}\, \int_0^{2\pi}\, \, 
\frac{\alpha(e^{i\phi}-1)}
     {\{(1-\alpha^2e^{i\phi})\, (1\, -\alpha^2e^{-i\phi})\}^{1/2}}
     \cdot \,  e^{i|m|\phi} \cdot \,  d\phi. 
\end{eqnarray}
We may reduce $\, a_{2m}$ to a hypergeometric function as
\begin{eqnarray}
\label{a2mfinal}
\hspace{-0.78in}
&&a_{2m} \,= \, \,
(1 \, -\alpha^2) \cdot \, \alpha^{2|m|} \cdot \, \frac{\Gamma(|m|+1/2)}{\pi^{1/2}|m|!}
\cdot  \,
{}_2F_1\Bigl([|m|+\frac{1}{2},\frac{1}{2}], \, [|m|+1], \, \alpha^4\Bigr), 
\end{eqnarray} 
which may be rewritten in terms of $\, k$ 
using (5) of page 111 of~\cite{batemanv1}:
\begin{eqnarray}
 &&\hspace{-.4in} \quad 
{}_2F_1\Bigl([m+\frac{1}{2},m+\frac{1}{2}], \, [,2m+1], \,  \, k^2\Bigr)
\nonumber \\
\label{F=F}
&&\hspace{-.4in} \quad  \quad  \quad
\, = \, \,\, (1+\alpha^2)^{2m+1} \cdot  \,
{}_2F_1\Bigl([m+\frac{1}{2},\frac{1}{2}], \, [m+1], \,  \, \alpha^4\Bigr). 
\end{eqnarray}
We find that: 
\begin{eqnarray}
&&\hspace{-1in}\, \,  a_{2m} \, =\,\,
\left(\frac{\alpha}{1+\alpha^2}\right)^{2|m|}
 \,\frac{1-\alpha^2}{1+\alpha^2}
\cdot \,\frac{\Gamma(|m|+1/2)}{\pi^{1/2}|m|!}
\cdot \,
{}_2F_1\Bigl([|m|+\frac{1}{2},|m|+\frac{1}{2}], \, [2\, |m| +1], \,\, k^2\Bigr)
\nonumber\\
\label{aeven}
&&\hspace{-.98in}
\,\, = \,\,\,
\Bigl({{k} \over {2}}\Bigr)^{2|m|} \cdot \, \sqrt{1-k^2}
\cdot \, \frac{\Gamma(|m|+1/2)}{\pi^{1/2}|m|!} \cdot \,
{}_2F_1\Bigl([|m|+\frac{1}{2},|m|+\frac{1}{2}],\, [2\, |m|+1], \, \, k^2\Bigr). 
\end{eqnarray}
Similarly:
\begin{eqnarray}
&&\hspace{-1.0in}
a_{2m+1} \, = \,\,
\left(\frac{k}{2}\right)^{2m+1} \, \frac{\Gamma(m+1/2)}{\pi^{1/2}m!} \cdot \,
\Bigl[\left(\frac{k}{2}\right)^2  \,\frac{m+1/2}{m+1} \cdot \,
 {}_2F_1\Bigl([m+\frac{3}{2},m+\frac{3}{2}], \, [2m+3], \, k^2\Bigr)
\nonumber\\
\label{aoddint1}
&& \quad  -\, {}_2F_1\Bigl([m+\frac{1}{2},m+\frac{1}{2}], \, [2m+1], \, \, k^2\Bigr)\Bigr]. 
\end{eqnarray} 
The two hypergeometric
functions in (\ref{aoddint1}) combine and thus, with the symmetry
(\ref{anmn1}), give the final result: 
\begin{eqnarray}
\label{aodd}
&&\hspace{-.9in} \, \,
a_{\pm (2m+1)} \, \, \, = \, \,
\nonumber \\
&&\hspace{-.9in} \, \, \quad 
 \mp \left(\frac{k}{2}\right)^{2|m|+1}
 \cdot \, \frac{\Gamma(|m|+1/2)}{\pi^{1/2}|m|!} \,
\cdot \, {}_2F_1\Bigl([|m|+\frac{1}{2},|m|+\frac{1}{2}], \, [2\, |m|+2],  \, \, k^2\Bigr). 
\end{eqnarray}

\vskip .1cm

\subsection{$C(0,N) \, $ for $\,T >\,T_c \, $ at $\,\nu\, =\,\,-k_>\,= \,\, -1/k$}
\label{subCONnu}
In the following we will simply denote $\, {\tilde K}(k)$, $\, {\tilde E}(k)$
and $\, {\tilde \Pi}(-k\nu,k)$ (see (\ref{Kdef}), (\ref{Edef}), (\ref{Pidef})) 
by $\, {\tilde K}$, $\, {\tilde E}$ and $\, {\tilde \Pi}$.
For $\,T\,>\,T_c\, $ we find, from (\ref{kalpha}) and (\ref{nualpha}),
that if $\, \nu\,= \,\,-k_>\,= \, \, -1/k\, $ then
\begin{equation}
\label{tgtc}
\hspace{-.01in} \quad 
\alpha_1 \,= \,\, -\alpha_2^{-1} \, =\,\,-\alpha,
\quad \, {\rm and}\, \quad \quad   k_> \, =\,\, \frac{2 \alpha}{\alpha^2+1},
 \quad \, s_h \, = \, -ik_>, \quad \, s_v\, = \, i, 
\end{equation}
and we find the matrix elements (\ref{an}) reduce to
\begin{equation}
\label{anpspecial}
\hspace{-.05in}
a_n \, = \, \, -\frac{1}{2\pi} \, \int_0^{2\pi} \, 
\, \left[\frac{1\, -\alpha^2\, e^{2i\theta}}{1\, -\alpha^2e^{-2i\theta}}\right]^{1/2}
\cdot \,  e^{(n-1)i\theta} \cdot \, d\theta. 
\end{equation}
By sending $\,\, \theta \, \rightarrow \, \, \theta\, +\pi\,\, $ we see that
$ \, \, a_n\, =\,\,  (-1)^{n-1}\, a_n$,
and thus
\begin{equation}
\label{aevenp}
\hspace{-.08in}
a_{2n}\,\, =\, \,\, 0, 
\end{equation}
and
\begin{eqnarray}
&&\hspace{-.2in}
a_{2n+1} \, \, = \,  \, \,
-\frac{1}{2\pi} \, \int_0^{2\pi} \, 
\left[\frac{1 \, -\alpha^2 \, e^{2i\theta}}{1 \, -\alpha^2 \, e^{-2i\theta}}\right]^{1/2}
\cdot \,  e^{2ni\theta} \cdot \,  d\theta 
\nonumber\\
\label{anoddp}
&&\hspace{.3in}
= \, \, \,  -\frac{1}{2\pi} \, \int_0^{2\pi} \, 
\, \left[\frac{1-\alpha^2e^{i\phi}}{1-\alpha^2e^{-i\phi}}\right]^{1/2}
\cdot \,   e^{ni\phi} \cdot \, d\phi, 
\end{eqnarray}
which we recognize as the matrix elements $\, a_{-n}$ of the diagonal
correlation (\ref{an}) for $\, T\,<\,T_c\,$ with $\,\alpha_1 \,=\,\,0 \, \,$
and $\,\,\alpha_2 \, \rightarrow\,\, \alpha^2$.

We further recognize  because of (\ref{aevenp}) that 
\begin{equation}
\label{c02np1}
\hspace{-1in}
C(0,2N+1)\,\, =\,\,\, 0, 
\end{equation}
and that the $\, \, 2N \, \times\,  2N\, $ determinants for $ \, C(0,2N)$
factorize as:
\begin{eqnarray}
\label{factordet}
&&\hspace{-.98in}
C(0,2N)\,\, = \,\,
\\
&&\hspace{-.9in}  \quad  \quad 
\begin{array}{|llll|}
{a}_{-1}&{a}_{1}&\cdots&{a}_{2N-3}\\
{a}_{-3}&{ a}_{-1}&\cdots&{a}_{2N-5}\\
\vdots&\vdots&&\vdots \\
{a}_{-(2N-1)}&{a}_{-(2N-3)}&\cdots&{a}_{-1}
\end{array} 
\times
\begin{array}{|llll|}
{a}_{1}&{a}_{3}&\cdots&{a}_{2N-1} \\
{a}_{-1}&{a}_{1}& \cdots&{a}_{2N-3} \\
\vdots & \vdots&& \vdots \\
{a}_{-(2N-3)}&{a}_{-(2N-5)}&\cdots&{a}_{1}
\end{array}  \, .
\nonumber 
\end{eqnarray}
For $\, \, 2n+1\,> \,0\, $   we express (\ref{anoddp}) in terms
of hypergeometric functions as
\begin{equation}
\label{anp}
\hspace{-.28in}
a_{2n+1} \,\, = \,\,\,
\alpha^{2n} \cdot \,\frac{\Gamma(n+1/2)}{{\sqrt \pi}n!} \cdot \,
{}_2F_1\Bigl([-\frac{1}{2}, n+\frac{1}{2}], [n+1], \, \, \alpha^4\Bigr), 
\end{equation}
and:
\begin{equation}
\label{anm}
a_{-(2n+1)}\,\, =\,\,\,
\alpha^{2n+2} \cdot  \, \frac{\Gamma(n+1/2)}{2\,{\sqrt \pi}(n+1)!}
\cdot \,
      {}_2F_1\Bigl([\frac{1}{2},n+\frac{1}{2}], \, [n+2], \,\,\alpha^4\Bigr). 
\end{equation}
As a special case we note  from (56) of~\cite{mjm} by use of
(\ref{tgtc}) for $\,\nu\,=\, -k_> \, $ that
\begin{eqnarray}
\hspace{-.28in}
&&C(0,2) \,\, = \,\,\,
k_{>}^{-2} \cdot\, \Bigl( {\tilde E}^2 \,\, -(1\,-k_{>}^2) \cdot \, {\tilde K}^2 \Bigr)
\nonumber\\
\hspace{-.28in}
&&\hspace{.41in}
=\, \,\, k_{>}^{-2} \cdot \,
\Bigl({\tilde E}\, \, -\sqrt{1 \,-k_>^2} \cdot \, {\tilde K}\Bigr) 
\cdot \,
\Bigl({\tilde E}\,\,  +\sqrt{1\,  -k_>^2} \cdot \, {\tilde K}\Bigr), 
\label{c02p}
\end{eqnarray}
which illustrates the factorization property of $\, C(0,2N)$. 
For small $\, k$ we have
\begin{eqnarray}
&&\hspace{-.48in}
C(0,2)\, \,= \,\,\, \frac{1}{8}\, \,k_>^2\,\,\,\, +\frac{1}{16}\,\, k_>^4
\,\,\, +\frac{39}{1024}\,\, k_>^6 \, \,
+\frac{53}{2048}\,\, k_>^8\, \,\, +\frac{1235}{65536}\,\, k_>^{10}
\nonumber\\
&&\hspace{.28in}
+\frac{1887}{131072\, }\,\, k_>^{12}\, \, \,
+\frac{382291}{33554432}\,\,  k_>^{14}\,\,\,
+ O(k^{16}), 
\end{eqnarray}
which using (\ref{tgtc}) is rewritten in terms of $\, \alpha \, $ as
\begin{equation}
\label{C02alpha}
\hspace{-.38in}
C(0,2) \,  \,=\, \, \,
\frac{1}{2}\,\alpha^2 \,\, -\frac{1}{16}\, \alpha^6
\,\, -\frac{1}{64}\, \alpha^{10} \, \,
-\frac{13}{2048}\, \alpha^{14} \, \,\, +O(\alpha^{18}).
\end{equation}
Using maple we find
\begin{eqnarray}
&&{\tilde E}\,\,\, +\sqrt{1-k_>^2} \cdot \,{\tilde K}
\,\,\, =\,\,\,\, \frac{2}{1 \, +\alpha^2} \cdot\, a_{1}
\\
&& {{1} \over {k_>^2 }} \cdot \, \Bigl({\tilde E}
  \,\,\, -\sqrt{1-k_>^2} \cdot \,  {\tilde K}\Bigr)
\,\,\,  = \,\, \,\, \frac{1\, +\alpha^2}{2} \cdot \, a_{-1}, 
\end{eqnarray}
or equivalently using 
\begin{equation}
\hspace{-.58in}
\alpha  \, \, = \, \,  \,  \frac{1\, -\sqrt{1 -k_>^2}}{k_>}, 
\end{equation}
we have:
\begin{eqnarray}
\label{second}
&& a_1 \, \,= \, \,\,
\frac{1-\sqrt{1 -k_>^2}}{k_>^2}\,  \cdot \,
\Bigl({\tilde E}\, \,   +\sqrt{1 -k^2} \cdot \, {\tilde K}\Bigr), 
\\
\label{first}
&&a_{-1} \,\, = \,\,\,
\frac{1\,+\sqrt{1 -k_>^2}}{k_>^2} \, \cdot \, 
\Bigl({\tilde  E} \, \, -\sqrt{1\,-k_>^2}\cdot \,{\tilde K}\Bigr). 
\end{eqnarray}
To generalize and derive (\ref{second}) and (\ref{first}) we treat
$\, a_{2n+1}/\alpha \,$ and $\, \alpha \, a_{-(2n+1)} \,$ separately.
These calculations are detailed in \ref{appFirst}.

\vskip .1cm

\subsection{Quadratic difference equations for $ \, C(M,N)$}
\label{quadra}
  
In general for $ \, M \, \neq \,  0$ the correlation $\, C(M,N)$
with $\, M < \, N$ can be written as an $\, N \, \times \,  N$ determinant
which is {\em not} Toeplitz. We will not use this  determinant representation
but, instead, use
quadratic difference equations~\cite{mccoy1,perk,mccoy2,mccoy3}
which relate the (high-temperature) 
correlation functions $ \, C(M,N)$ for $ \, T \, > \, T_c\, $ to
the {\em dual correlation} $\, C_d(M,N)$ for $ \, T \,> \, T_c$, 
defined as the low temperature correlation with the replacement:
$ \, s_v  \,\rightarrow \quad  1/ s_h \, \hbox{ and} \,  \,  
s_h  \, \rightarrow \, 1/ s_v$
\begin{eqnarray}
&& \hspace{-.68in}
  s^2_h \cdot \,[C_d(M,N)^2 \, -C_d(M,N-1) \cdot \, C_d(M.N+1)]
\nonumber \\
&& \hspace{-.18in}
+[C(M,N)^2\, -C(M-1,N) \cdot \, C(M+1,N)] \,\, =\,\,\,  0,
\\
&& \hspace{-.68in} s^2_v \cdot \,[C_d(M,N)^2\, -C_d(M-1,N) \cdot \, C_d(M+1,N)]
\nonumber \\
&&\hspace{-.18in}
+[C(M,N)^2\, -C(M,N-1) \cdot \, C(M,N+1)] \,\, = \,\,  \,0, 
\\
&& \hspace{-.68in} s_v s_h \cdot \,
[C_d(M,N) \cdot \, C_d(M+1,N+1)\, -C_d(M,N+1) \cdot \, C_d(M+1,N)]
\nonumber \\
&&\hspace{-.18in}
= \, \,  C(M,N) \cdot \, C(M+1,N+1)\, -C(M,N+1) \cdot \, C(M+1,N), 
\end{eqnarray}
which hold for all $\,M$ and $\,N$, except 
$\,M =\,\,0,\,\, N =\,\, 0$,
where we have:
\begin{eqnarray}
&&C(1,0)\, \,  = \,\,\,  (1+ s^2_h)^{1/2} \,\,\,  -s_h \cdot \, C_d(0,1),
\\
&&C(0,1)\,\,  =\,\,\,  (1+ s^2_v)^{1/2} \,\,\,  - s_v \cdot \,C_d(1,0).
\end{eqnarray}
with $\, s_h\,=\,\sinh 2E_h/kT\, $ and $\,\, s_v \,=\,\, \sinh 2E_v/kT$.

From these quadratic difference equations we find~\cite{mjm}
for example for $\, T\, < \, T_c\, $ where $\, k\, = \,\, (s_vs_h)^{-1}$:
\begin{eqnarray}
&&\hspace{-.78in}
C(1,2) \,  = \, \, 
 s_v^2 \cdot \, (s_v^{-2}+1)^{1/2} \cdot \, \Bigl(
 \, s_h^{-2} \cdot \,   (s_v^{-2}s_h^{-2} -1) \cdot  \, 
{\tilde K}^2 \, \,    + (s_h^{-2} -1) \cdot \,  
{\tilde E} \, {\tilde K} \, \, + E^2
\nonumber \\
&&\hspace{-.18in}
+(s_v^{-2}-1) \, ( s_h^{-2}+1) \cdot \,  
 {\tilde E}\,  {\tilde \Pi} 
\,  \, \,   - (s_h^{-2}+1) \cdot \, (s_v^{-2}s_h^{-2}-1)   
\cdot \,  {\tilde K} \,   {\tilde \Pi} \Bigr), 
\end{eqnarray}
and for $\, T\, > \,  T_c$ where $\, k_>\,  = \,  \,  s_v\, s_h$:
\begin{eqnarray}
\label{C12bis}
&&\hspace{-.78in}
C(1,2) \,  =\,  \, 
\frac{(s_v^2 +1)^{1/2}}{s_h^2s_v} \cdot \, 
\Bigl( {\tilde E}^2 \, \, -(s_h^2\,s_v^2 -1) \cdot \, {\tilde K}^2  \, \,
+ (s_h^2\,s_v^2\,  +s_v^2 \, -2 ) \cdot \,  {\tilde E} \,  {\tilde K}
\nonumber \\
&& +(s_h^2 \, +1) \cdot \,  (s_v^2 \, -1)
 \cdot \,  {\tilde E} \,  {\tilde \Pi}  \,   \, \,  
+ (s_h^2\, \, +1) \cdot \,  (s_h^2 s_v^2 \,  -1)
\cdot  \,  {\tilde K} \,  {\tilde \Pi} \Bigr).
\end{eqnarray}
For $\, T\, < \, T_c\,$ and $ \, \nu\, =\,  -k$,  where
$\, s_h\, = \,\,  i, \, \, s_v\, =\,\,  -i/k$,  one has:
\begin{eqnarray}
\hspace{-.68in} \quad 
C(1,2)\, \, = \, \,\,  -(1 \, -k^2)^{1/2} \cdot \, k^{-2} \cdot \,
\Bigl( (1 \, -k^2) \cdot  \,  {\tilde K}^2  \,\,
   -2 \cdot \,  {\tilde E} \,  {\tilde K}\,\, \,   +{\tilde E}^2\Bigr). 
\end{eqnarray}
For $\, T\, > \, T_c \, $ and $\, \nu\, =\, -k_> \, $ where
$\, s_h\, =\, -i\, k_>, \, \, s_v=\, \, i\,\, $ one gets:
\begin{eqnarray}
\hspace{-.08in} \quad \quad \quad \quad 
C(1,2) \, \, = \,\,  \, 0.
\end{eqnarray}
Further special cases are given in \ref{appA}.

\vskip .1cm 

\section{Two-parameter family of nonlinear differential equations for $\, C(M,N)\, $ for $\nu\, =\,\,   -k$}
\label{nonlin}

We have obtained a nonlinear equation with the Painlev{\'e} property
(i.e. fixed critical points)
which is satisfied by $\, C(M,N)$ for $\, \nu\,  =\, -k\, $ by using the program
{\it  guessfunc} developed by J. Pantone~\cite{Pantone}. This program searches for
nonlinear equations satisfied by series expansions. We have applied
this program for many values of the integers $\, M,\, N$
for the series expansions of $\, C(M,N)$
at $\, \nu\, =\, -k \, $ obtained from either the Toeplitz determinants for
$\, C(0,N)$ of section 2, or from expressions deduced from the
quadratic recursion relations of section \ref{quadra}. The results are as follows.

\vskip .1cm

\subsection{Nonlinear differential equations for $\, C(M,N)$ for $\nu\, =\,\,   -k$ and
    $\, M \le N$: the low-temperature case }
\label{nonlinlow}

For $\, T\, <\, T_c \, $ and $\, \nu\, =\, -k \, $ with $\, t=\, k^2\, $ and
\begin{eqnarray}
\hspace{-.08in}
\sigma\,\, =\,\, \,
t \cdot \, (t-1) \cdot \, \frac{d \ln C(M,N)}{dt} \,\, \,-\frac{t}{4}, 
\end{eqnarray}
we have:
\begin{eqnarray}
&&\hspace{-.38in}
  [t \cdot \, (t-1) \cdot \, \sigma'']^2 \, \, 
  +4 \cdot \,  \{ \sigma' \cdot \, (t\, \sigma'\,  -\sigma) \cdot \,
  ((t-1)\, \sigma'\,  -\sigma) \}
\nonumber\\
&&\hspace{-.08in}
-M^2 \cdot \, (t\, \sigma'-\sigma)^2 \,\, -N^2 \cdot \, \sigma'^2
\nonumber\\
\label{eqnm}
&&\hspace{-.08in}
+[M^2 +N^2 \,
  -\frac{1}{2} \, (1 \, +(-1)^{M+N})]   \cdot \,  \sigma' \cdot \, (t\, \sigma'\, -\sigma)
\, \,  = \, \, \, 0. 
\end{eqnarray}
Note that when $\, M = \, N$ the diagonal correlation $ \, C(N,N)$ {\em does not
  depend on the anisotropy variable} $\, \nu$. There is no difference between
the diagonal correlation functions $ \, C(N,N)$ for $\, \nu \, =\, -k \, $ and
for arbitrary $\, \nu$. As expected the two-parameters $ \, (M, \, N)$-family of
nonlinear differential equations (\ref{eqnm}) actually reduces 
when $\, M \,=\, N$ to the Jimbo-Miwa
nonlinear differential equation  (\ref{jmequation}) for the diagonal
correlation $ \, C(N,N)$ for $\, T\, < \, T_c$.

\vskip .1cm

\subsection{Nonlinear differential equations for $\, C(M,N)\, $ for
  $\nu\, =\,\,   -k$, $\, M \le \, N \, $ and $\,M+N$ even: the high-temperature case}
\label{nonlinhigh}
 
For $\, T\, >\, T_c\, $ and $\, \nu=\, -k_> \, $ with $\, M \le N \, $
and $\, M+N$ {\em even} with  $\, t=\, k_>^2 \, $ and
\begin{eqnarray}
  \sigma\,\, = \,\,\,\,
  t\cdot \, (t-1) \cdot \, \frac{d \ln C(M,N)}{dt}\,\,\, -\frac{1}{4}, 
\end{eqnarray}
we have: 
\begin{eqnarray}
\label{EqMNhigh}
  &&\hspace{-.38in}
  [t \cdot \, (t-1) \cdot \, \sigma'']^2 \, \, 
  +4 \cdot \, \{\sigma' \cdot \, (t\, \sigma'\,  -\sigma)
  \cdot \, ((t-1) \, \sigma'\,  -\sigma)\}
\nonumber\\
&&\hspace{-.08in}
-M^2 \cdot \, (t\, \sigma' \, -\sigma)^2 \, \,\,
+(N^2+M^2\, -1) \cdot \, \sigma' \cdot \, (t\, \sigma' \, -\sigma)
\nonumber\\
&&\hspace{-.08in}
-N^2 \cdot \, \sigma'^2 \,\,\,
-\frac{1}{4}\,  (N^2-M^2) \cdot \, (t\, \sigma' \, -\sigma)
\nonumber\\
&&-\frac{1}{4} \cdot \, (N^2-M^2) \cdot \, \sigma'
\, -\frac{1}{16}  \cdot \, (N^2 -M^2)^2
\, \,= \, \,\,  0. 
\end{eqnarray}
As expected, when $\, M \,=\, N$, the two-parameters $ \, (M, \, N)$-family of
nonlinear differential equations (\ref{EqMNhigh})  {\em also reduces 
to the Jimbo-Miwa nonlinear differential equation}  (\ref{jmequation}) for the diagonal
correlation $ \, C(N,N)\, $ for $\, T\, > \, T_c$.

\vskip .1cm

\subsection{Nonlinear differential equation for $C(M,N)\, $ for
  $\nu= \, -k_>$, $\, M < \, N \, $ and $ M+N$ odd: the high temperature case}
\label{subsec}

For $\,  T>\,  T_c$ and $\nu=\,  -k_>\, $ we found in (\ref{c02np1}) that
$\,  C(0,2N+1)\,  =\,  0$. For $\,  C(0,1)$ this vanishing occurs because of
the vanishing of the factor $\,  \sqrt{1\,  +\nu/k_>}\, $ in (\ref{c01p}) 
and we found, for instance in (\ref{c01tp}), that:
\begin{equation}
 \hspace{-.98in}
 \lim_{\nu\rightarrow -k_>} \,
 \Bigl(1\,  +{{\nu} \over {k_>}}\Bigr)^{-1/2} \cdot \,   C(0,1)
\,\,   = \,\, \,  \,    {{ {\tilde K} \, -{\tilde E}} \over {k_>}}.  
\end{equation}
We have examined this phenomenon more generally by considering
$\,  C(M,N)$ for low values of $\, M, \, N$ and find that 
for $\, M+N$ odd 
\begin{equation}
\hspace{-.68in}
C(M,N)\, \, =\, \, \, 0, 
\end{equation}
and that the limit
\begin{equation}
\label{tildec}
\hspace{-.98in}
\lim_{\nu\rightarrow -k_>} \,
\Bigl(1 \, +{{\nu} \over {k_>}} \Bigr)^{-1/2} \cdot \, C(M,N)
\,\, \, = \, \,\,  \, {\tilde C}(M,N), 
\end{equation}
exists and is nonzero. For example
\begin{eqnarray}
&&\hspace{-.7in} \quad \quad 
{\tilde C}(0,3) \,  \, = \, \,  \,
 \frac{4}{k_>^5} \cdot \,  \Bigl[(k_>^2-1)^2 \cdot \,  {\tilde K}^3 \, \, 
 -(2k_>^2 -3) \cdot \, (k_>^2 -1) \cdot \,  {\tilde K}^2\,  {\tilde E}
\nonumber \\
&& \hspace{-.1in} \quad  \quad  \quad 
-(2k_>^2-3) \cdot \,  {\tilde K}\,  {\tilde E}^2
     \, \,  \,  -(k_>^2+1) \cdot \,  {\tilde E}^3 \Bigr].
\end{eqnarray}
We define
\begin{equation}
  \sigma\, \, =\, \, \,
  t \cdot \, (t-1) \cdot \, \frac{d\ln {\tilde C}(M,N)}{dt}
       \,\, \,  \,-\frac{1}{4}, 
\end{equation}
and find:
\begin{eqnarray}
&&\hspace{-.6in}
  [t \cdot \, (t-1) \cdot \, \sigma'']^2 \, \, 
  +4 \cdot \, \{\sigma' \cdot \, (t\, \sigma'\,  -\sigma)
  \cdot \, ((t-1) \, \sigma'\,  -\sigma)\}
\nonumber\\
&&\hspace{-.3in}
-M^2 \cdot \, (t \, \sigma' \, -\sigma)^2 \, \,\,
+(N^2+M^2\, -2) \cdot \, \sigma' \cdot \, (t\, \sigma' \, -\sigma)
\nonumber\\
&&\hspace{-.3in}
-N^2 \cdot \, \sigma'^2 \, \, \,
-\frac{1}{4} \cdot  \,  (N^2-M^2-1) \cdot \, (t\, \sigma' \, -\sigma)
\, \, \, -\frac{1}{4} \cdot \, (N^2-M^2+1) \cdot \, \sigma'
\nonumber\\
\label{eqnp}
&&\hspace{-.3in}
 -\frac{1}{16} \cdot \, (N^2 -M^2)^2 \,\,\,  +\frac{1}{8}\cdot \, (M^2+N^2-1)
\, \,\, = \, \,\,  0. 
\end{eqnarray}

\vskip .1cm

{\bf Remark:} All the previous 
sigma non-linear ODEs  (\ref{eqnm}), (\ref{EqMNhigh}) and (\ref{eqnp})
are valid for $\, \nu \, = \, -k \, $ and $\, M\, \le N$. However recalling
the symmetry relations (\ref{cmnsym}) and especially (\ref{mnsym}),
$\, C(M, \, N; \, \nu=-k)\, \,=\, \, C(N, \, M; \, \nu= -1/k)$, it is
straightforward to see that these sigma non-linear ODEs (\ref{eqnm}),
(\ref{EqMNhigh}) and (\ref{eqnp})
{\em are also valid} for
the $\, C(M, \, N)$ correlation functions  for
$\, M \, \ge \, N$ {\em but when} $\, \nu \, = \,\, -1/k$.

\vskip .1cm

\subsubsection{A Kramers-Wannier formal symmetry. \\}
\label{intriging}

In~\cite{Holonomy} a representation of the Kramers-Wannier duality
on $\, \sigma$ has been introduced\footnote[1]{See equation (16) page 78 
of~\cite{Holonomy}.}:
\begin{eqnarray}
\label{dual}
\hspace{-0.98in}&&  \quad \quad \quad  \quad  \,  \,  \,  
(t, \, \sigma, \, \sigma', \,   \sigma'') \quad\quad \longrightarrow 
\quad  \quad \,
\Bigl({{1} \over {t}}, \, \,  {{\sigma } \over {t}}, \,\,  
\sigma \, - \, t \cdot \,\sigma',
\,  \, t^3 \cdot \, \sigma''   \Bigr).                
\end{eqnarray}
It had been noticed that this involutive transformation (\ref{dual})
preserves the sigma form of Painlev\'e VI equation (\ref{jmequation}).
This transformation just amounts to saying, for any function $\, F(t)$,
that the change of variable $\, \, t \, \rightarrow \, \, 1/t \, $ changes 
\begin{eqnarray}
  \quad 
  \sigma \,\,  =\,\, \, \, 
  t \cdot \, (t-1) \cdot \, \frac{d \ln(F(t))}{dt}\,\, \, -\frac{1}{4}, 
\end{eqnarray}
into $\, \, \tilde{\sigma}/t$ where:
\begin{eqnarray}
  \quad 
  \tilde{\sigma} \,\,  =\,\, \,
  t \cdot \, (t-1) \cdot \, \frac{d \ln(F(t))}{dt}\,\, \, -\frac{t}{4}. 
\end{eqnarray}
and vice-versa. 

$\bullet$ One first remarks that this involutive transformation (\ref{dual}) 
{\em actually transforms the (low-temperature) nonlinear differential equation}
(\ref{eqnm}) {\em into itself where  $\, M$ and $\, N$ are permuted}.

$\bullet$ One then remarks that this involutive transformation (\ref{dual}) 
{\em also transforms the (high-temperature, $\, M+N$ even)
  non-linear differential equation} (\ref{EqMNhigh})
{\em into itself where  $\, M$ and $\, N$ are permuted}.

$\bullet$ One finally remarks that this involutive transformation (\ref{dual}) 
{\em also transforms the (high-temperature, $\, M+N$ odd) non-linear differential equation}
(\ref{eqnp}) {\em into itself where  $\, M$ and $\, N$ are permuted}.

These three results must be seen as mathematical symmetries: the
question of the physical interpretation of the non-linear differential equations
(\ref{eqnm}), (\ref{EqMNhigh}) and (\ref{eqnp}) when  $\, M$ and $\, N$ are permuted,
remains an open question. 

Note that $\, \nu$ {\em is left invariant by the Kramers-Wannier
duality}\footnote[5]{The Kramers-Wannier
duality changes $s_h \, \rightarrow \, s_v^{*} \, = \, \, 1/s_v$,
  $s_v \, \rightarrow \, s_h^{*} \, = \, \, 1/s_h$
and thus $\,s_h/s_v \, \rightarrow \,s_h/s_v$. },
in contrast with $\, k$ which becomes its reciprocal $\, k \,  \rightarrow \, 1/k$.
Consequently, the selected condition $\, \nu \, = \, -k$ is
{\em not left invariant by the Kramers-Wannier duality}. 
The high-temperature non-linear differential equations
(\ref{EqMNhigh}) and (\ref{eqnp}),
valid at $\, \nu \, = \, -k \, $ have no reason to be deduced
from the low-temperature non-linear
differential equations (\ref{eqnm}), valid at $\, \nu \, = \, -k$,
using a Kramers-Wannier-like duality (\ref{dual}).

Along this line, let us note, for $\, M+N$ even,  that one can change the 
low-temperature non-linear differential equation  (\ref{eqnm}) into
the high-temperature  non-linear differential equation  (\ref{EqMNhigh})
using the involutive transformation:
\begin{eqnarray}
\label{invol}
\hspace{-0.98in}&& \quad \quad  \quad 
(\sigma, \, \sigma', \,   \sigma'',\, \,  M, \, \, N) \quad \quad \longrightarrow 
\nonumber \\
\hspace{-0.98in}&&  \quad 
\quad \quad
% \longrightarrow 
\quad  \quad \, \quad 
\Bigl(\sigma \, + \, {{N^2 \, -M^2 } \over {4}} \cdot \, (t \, -1), \,\,\,
\sigma' \, + \, {{N^2 \, -M^2 } \over {4}}, \,  \,\sigma'',\,\,   N, \, \, M
\Bigr).                
\end{eqnarray}

\vskip .1cm

\section{Sigma form of Painlev\'e VI: Okamoto parameters}
\label{Oka}

The search for nonlinear differential equations with the Painlev{\'e}
property is an ongoing field of research~\cite{Cosgrove} and is far from being 
complete even for equations of second order. However for equations
of the form
\begin{equation}
\label{yF}
\hspace{-.98in} (y'')^2 \,\,  = \,\, \, F(y,\,y',\,x), 
\end{equation} 
with {\em fixed singularities} at $\,x= \,0,\,1,\,\infty$, 
a  solution was given by Cosgrove and Scoufis in (4.9)
of~\cite{cosgrove} where 
it is shown that the non-linear differential equation with six parameters
\begin{eqnarray}
&&\hspace{-.38in} \quad 
  (x \cdot \, (x-1) \cdot \, y'')^2 \, \,
  +4 \cdot \, \{y' \cdot \, (xy'-y)^2 \, - y'^2 \cdot \, (xy' \, -y)
\nonumber\\
&&\hspace{-.08in} \quad
+c_5 \cdot \, (xy'-y)^2 \,\, 
+c_6 \cdot \, y' \cdot \, (xy'-y) \, \, +c_7 \cdot \, (y')^2
\nonumber\\
\label{cosgroveeq}
&&\hspace{-.08in} \quad 
+c_8 \cdot \, (xy'-y) \, \, \,  +c_9\cdot \, y'\,  \, \, +c_{10}\}
\, \, = \,\, \,  0, 
\end{eqnarray} 
has the Painlev{\'e} property\footnote[1]{Namely has fixed critical points.}
and is birationally equivalent to
Painlev{\'e} VI. Both equations (\ref{eqnm}) and (\ref{eqnp}) are of
the form (\ref{cosgroveeq}) and hence are sigma forms of Painlev{\'e} VI. 
The non-linear differential equation (\ref{cosgroveeq}) is invariant in form
under the transformation
\begin{equation}
\label{cosgrovesym}
\hspace{-.68in}
y \,\,  = \, \,\,  {\bar y} \, \,  \, +A \cdot \,x \,+B, 
\end{equation}
which transforms the six parameters $\, c_k$ into new parameters
$ \,{\tilde  c}_k$ as follows
\begin{eqnarray}
\label{c6t}
&&\hspace{-.8in}
{\tilde c}_5 \, = \,\, c_5 \,+A, \quad \quad  \quad \quad 
{\tilde c}_6\, = \,\, c_6 \,-2B \,-2A, 
\\
\label{c7t}
&&\hspace{-.8in}
{\tilde c}_7\, = \,\, c_7 \, +B,  \quad \quad  \quad \quad 
{\tilde c}_8 \, = \,\, c_8 \, -2 AB \, -A^2 \, -2B \cdot \, c_5 \, +A \cdot c_6, 
\\
\label{c9t}
&&\hspace{-.8in}
{\tilde c}_9\,  = \,\,  c_9 \, +B^2 \, +2 AB \,-B \cdot  c_6 \, +2A \cdot  c_7, 
\\
\label{c10t}  
&&\hspace{-.8in}
{\tilde c}_{10}\, = \,\,
c_{10} \, +AB^2 \, +A^2B \, +B^2 \cdot  c_5 \,
  -AB \cdot c_6 \, +A^2 \cdot c_7 \, -B \cdot  c_8 \, +A \cdot c_9.
\end{eqnarray}
The canonical form  of sigma Painlev{\'e} VI given by Okamoto~\cite{oka} 
which depends on four parameters $ \,n_1, \,n_2, \,n_3, \, n_4$ reads
\begin{eqnarray}
&&\hspace{-.48in}
  h' \cdot \, \{t  \cdot \, (t-1) \cdot  \,h''\}^2
      \,  \, +\{h' \cdot \,(2h \,-(2t-1) \,h') \,\, +n_1n_2n_3n_4 \}^2
\nonumber\\
\label{okacanonical}
&&\hspace{-.38in} \quad \quad 
-(h' \,+n_1^2) \cdot \, (h'\, +n_2^2)\cdot \, (h'\, +n_3^2)\cdot \, (h'\, +n_4^2)
  \, \, \, = \, \,  \, 0, 
\end{eqnarray} 
which when expanded and removing the common factor of $ \, h'$
reads
\begin{eqnarray}
\label{Cosgrove}
&&\hspace{-.7in} \quad \quad 
\{t  \cdot \, (t -1) \cdot  \, h''\}^2 \, \,
+4 \, h' \cdot \, (t\, h' \, -h) \cdot \, ((t\, -1)\, h' \, -h)
\nonumber \\
&&\hspace{-.7in} \quad \quad \quad 
\, \, + 4 \, c_{10} \, \,\, + 4 \, c_{9}  \cdot \,h'
\, \,   \, + 4 \, c_{8} \cdot \, (t \cdot \,h' \, -h)\, \, \, 
\, \, + 4 \, c_{7}  \cdot \,h'^2 
\, \, \, = \, \,  \, 0, 
\end{eqnarray}
which is of the Cosgrove form (\ref{cosgroveeq}) with
\begin{eqnarray}
\label{oka7}
&&\hspace{-.7in}
c_7\,  = \,\,  -(n_1^2+n_2^2+n_3^2+n_4^2)/4, \quad  \quad  \quad 
c_8\,  =\, \,  -n_1n_2n_3n_4,  
\\
\label{oka9}
&&\hspace{-.7in}
c_9\, =\, \,  -(n_1^2n_2^2+n_1^2n_3^2 \, +n_1^2n_4^2+n_2^2n_3^2
\, +n_2^2n_4^2+n_3^2n_4^2 \, -2n_1n_2n_3n_4)/4, 
\\
\label{oka10}
&&\hspace{-.7in}
c_{10} \, = \, \, -(n_1^2n_2^2n_3^2 \, +n_1^2n_2^2n_4^2 \, +n_1^2n_3^2n_4^2
\, +n_2^2n_3^2n_4^2)/4.  
\end{eqnarray}
We  see from  (\ref{c6t}) if we choose
\begin{equation}
\label{linear}
\hspace{-.8in}
A \, =\, \, -c_5, \quad \quad \quad \quad   B \, = \, \,  {{c_6} \over {2}} \,\,  +c_5, 
\end{equation}
that 
\begin{equation}
\hspace{-.9in}
{\tilde c}_5 \,\, = \, \,\, {\tilde c}_6 \,\, = \,\, \, 0, 
\end{equation}
and thus the general  non-linear differential
equation (\ref{cosgroveeq}) is reduced to the
Okamoto form with the Okamoto parameters determined from
(\ref{oka7})-(\ref{oka10}).

\vskip .1cm

\subsection{Okamoto parameters for $\, T< \, T_c$ and $\, \nu\, =\, -k \, $}
\label{Okaparam}

For $ \, T < \, T_c\,$ and $\, \nu\, =\, -k \, $
we  obtain the Okamoto parameters for (\ref{eqnm})
with the  parameters which shift from (\ref{eqnm}) to the canonical
Okamoto form  determined from (\ref{linear}), to be
\begin{equation}
\label{linearm}
\hspace{-.6in} \quad \quad \quad
A \, = \,\, \frac{M^2}{4}, \quad \quad \quad
B \, =\, \, \frac{1}{8} \cdot \, \Bigl(N^2-M^2\, \, -{{1 \, +(-1)^{M+N}} \over {2}}\Bigr). 
\end{equation}
Thus we find from (\ref{c7t})-(\ref{c10t}) that
\begin{eqnarray}
&&\hspace{-1in}
  {\tilde c}_7\,  = \,\,
  -\frac{1}{8} \cdot \, \Bigl(N^2+M^2 \, +{{1 \, +(-1)^{M+N}} \over {2}}\Bigr), 
\quad \quad \quad 
      {\tilde c}_8\,  = \,\,
      \frac{M^2}{16} \cdot \, \Bigl(N^2\, -{{1 \, +(-1)^{M+N}} \over {2}}\Bigr), 
\nonumber  \\
&&\hspace{-1in}
{\tilde c}_9\, =\,\,
-\frac{1}{64} \cdot \, \Bigl(N^2 -M^2\, -{{1 \, +(-1)^{M+N}} \over {2}}\Bigr)^2
  \,  \, \, -\frac{N^2 M^2}{8}, 
\\
&&\hspace{-1in}
{\tilde c}_{10} \,
= \,  \, -\frac{M^2}{128} \cdot \,
\Bigl[ \Bigl(N^2-M^2\, \,  -{{1 \, +(-1)^{M+N}} \over {2}}\Bigr)
 \cdot \, \Bigl(N^2 \,\,  -{{1 \, +(-1)^{M+N}} \over {2}}\Bigr)  \,\,  +2\, M^2N^2\Bigr],
\nonumber 
\end{eqnarray}
and thus, from (\ref{oka7})-(\ref{oka10}), we obtain the Okamoto
parameters (unique up to permutations and the change of
an even number of signs)
\begin{eqnarray}
&&\hspace{-1in} \quad  \quad 
n_1\,=\,\,
\frac{1}{2}\cdot \, \Bigl(N\,- {{1+(-1)^{M+N}} \over {2}}\Bigr),
\quad \quad  \quad  n_2\, = \,\,
\frac{1}{2} \cdot \, \Bigl(N \, + {{1+(-1)^{M+N}} \over {2}}\Bigr),
\nonumber\\
\label{okam}
&&\hspace{-1in} \quad  \quad 
n_3\, =\,\, \frac{M}{2},  \quad \quad  \quad \quad  \quad    n_4\, =\, \, -\frac{M}{2}, 
\end{eqnarray}
with
\begin{eqnarray}
\label{shiftcmnm}
\hspace{-.4in} 
h \,\, =\, \,\,\, t \cdot \, (t-1) \cdot \,\frac{d\ln C(M,N)}{dt}
\nonumber \\
\hspace{-.4in} \quad  \quad \quad  \quad 
\,\,  -\frac{M^2+1}{4}\cdot \,t \,\, \,\,\, \,
-\frac{1}{8}  \cdot \,
\Bigl(N^2-M^2\,\, \,  -{{1 \, +(-1)^{M+N}} \over {2}}\Bigr), 
\end{eqnarray}
where $\, t= \, k^2$.

\vskip .1cm

\subsection{Okamoto parameters for $ \, T>\, T_c$, $\,\,\nu\,=\,\,-k_> \, $ and $\, M+N$ even}
\label{subOka}

For $\, T> \, T_c$,  $\,\,\nu\,=\,\,-k_> \, $ and $\, M+N$ {\em even}
we find for (\ref{EqMNhigh}), from (\ref{linear}), that:
\begin{equation}
\hspace{-.8in} 
A\,\,=\,\, \, \frac{M^2}{4}, \quad \quad \quad
B \,\, =  \, \,\, \frac{1}{8}\cdot \, (N^2-M^2-1).
\end{equation}.
Thus with $\,t =\,k_>^2 $
\begin{eqnarray}
\label{hevenp}  
\hspace{-.9in} \quad \quad 
h \, \, = \, \,\,\,  t \cdot \, (t-1) \cdot \,
\frac{d\ln  (C(M,N)}{dt}\,\,\,\, -\frac{M^2}{4} \cdot \, t \,\,\,\,\,
-\frac{1}{8} \cdot \, \Bigl(N^2-M^2\, \, +1\Bigr),
\end{eqnarray}
satisfies (\ref{okacanonical}) with
the Okamoto parameters: 
\begin{equation}
\label{okap}
\hspace{-1in}  \quad  \quad \quad \quad
n_1\, = \, \, \frac{M-1}{2}, \quad \quad
n_2\, =\,\,  \frac{M+1}{2}, \quad \quad
n_3\, =\, \, \frac{N}{2},  \quad \quad
n_4\, = \,\,  -\frac{N}{2}. 
\end{equation}

\vskip .1cm 

\subsection{Okamoto parameters for $ \, T > \, T_c$, $\,\nu\,=\,\,-k_>$, $\, M \, <\, N$ and $\, M+N$ odd}

For $\,  T> \, T_c\, $ we find for (\ref{eqnp}), from (\ref{linear}), that 
\begin{eqnarray}
\hspace{-.6in} \quad  \quad  \quad 
A\,\, =\, \,\, \frac{M^2}{4}, \quad \quad \quad
B\, \,= \, \, \,\frac{1}{8}\cdot \, (N^2-M^2-2). 
\end{eqnarray}.
Thus with $\, \, t= \, k_>^2$
\begin{eqnarray}
\label{htildep}  
\hspace{-.8in} \quad \quad \, \, \,  
h\, \, \,  =\, \, \,\, \,
t  \cdot\,  (t-1) \cdot \,  \frac{d}{dt}\ln {\tilde C}(M,N)
\,  \,\, \,  \,  -\frac{M^2}{4}\cdot \, t
\,\, \,\,   \,   \,   -\frac{1}{8} \cdot \, (N^2-M^2),  
\end{eqnarray}
satisfies the Okamoto equation (\ref{okacanonical}) with:
\begin{eqnarray}
\label{okapodd}
\hspace{-.8in}\quad
n_1\, \,=\, \,\, \frac{M-1}{2},\quad\, n_2 \, \,=\,\, \, \frac{M+1}{2},
\quad \, n_3\,\, =\, \,\, \frac{N+1}{2}, \quad \,
n_4\,\, = \,\, \,-\frac{N-1}{2}.
\end{eqnarray}

\vskip .1cm 

\subsection{Boundary conditions}
\label{boundary}

In order to complete the specification
$\,\nu\,=\,\,-k\, $  of $ \, C(M,N)$ we must specify the
boundary conditions for the nonlinear equations (\ref{eqnm}), (\ref{EqMNhigh})
and (\ref{eqnp}). This is most systematically  done by first determining
the allowed boundary conditions on the canonical equation of Okamoto
(\ref{okacanonical}) which are analytic at $ \, t= \, 0$.
A  detailed and explicit analysis of these boundary conditions
is performed in \ref{boundaryd}.

\vskip .1cm

\section{Relation to the determinants of Forrester-Witte}
\label{Forrester}

These results should be compared with the results of
Forrester-Witte~\cite{fw} as given in~\cite{gil}  as
\begin{equation}
\label{det}
\hspace{-.6in} 
D^{(p,p',\eta,\xi)}_N(t)
\,\, = \, \,\,  {\rm det}\left[A^{(p,p',\eta,\xi)}_{j-k}(t)\right]_{j,k=0}^{N-1}, 
\end{equation}
where
\begin{eqnarray}
&&\hspace{-.8in}  
  A_m^{(p,p',\eta,\xi)}(t) \,  \, = \, \,
\\
\label{aa1}
&&\hspace{-.6in}    \frac{\Gamma(1+p') \, t^{(\eta-m)/2} \, (1-t)^{p}}
{\Gamma(1+\eta-m) \, \Gamma(1-\eta+m+p')} \cdot \,
{}_2F_1\Bigl([-p,1+p'], \, [1+\eta-m], \, \, \frac{t}{t-1}\Bigr)
\nonumber\\
\label{a1}
&&\hspace{-.6in} \quad 
+\frac{\xi \cdot \, \Gamma(1+p)t^{(m-\eta)/2} \, (1-t)^{p'}}
{\Gamma(1-\eta+m) \, \Gamma(1+\eta-m+p)} \cdot \,
{}_2F_1\Bigl([-p', 1+p], \,[1-\eta+m], \, \, \frac{t}{t-1}\Bigr),
\nonumber
\end{eqnarray}
which, using the identity (6) on page 109 of~\cite{batemanv1}
\begin{equation}
\hspace{-.7in} 
 {}_2F_1([a,b], \,  [c], \,\, t) \, \, = \, \, \,
 (1-t)^{-a} \cdot \,
   {}_2F_1\Bigl([a,c-b],\, [c], \, \, \frac{t}{t-1}\Bigr), 
\end{equation}
is rewritten as:
\begin{eqnarray}
&&\hspace{-.8in}
A_m^{(p,p',\eta,\xi)}(t) \, \, \, = \, \,
\\
\label{aa2}
&&\hspace{-.6in}
\frac{\Gamma(1+p') \, t^{(\eta-m)/2}}
 {\Gamma(1+\eta-m) \, \Gamma(1-\eta+m+p')} \cdot \,
 {}_2F_1([-p,-p'+\eta-m], \, [1+\eta-m], \, \, t)
\nonumber\\
\label{a2}
&&\hspace{-.6in} \quad 
+\frac{\xi \cdot \, \Gamma(1+p) \, t^{(m-\eta)/2}}
{\Gamma(1-\eta+m) \, \Gamma(1+\eta-m+p)} \cdot \,
{}_2F_1([-p', -p-\eta+m], \, [1 -\eta+m], \, \, t).
\nonumber
\end{eqnarray}
 From (2,27) of~\cite{gil} 
\begin{eqnarray}
\label{htau}
&&\hspace{-.5in}
h \, \,  = \, \,  \,
t \cdot \, (t-1) \cdot \,
\frac{d}{dt}\ln\left(t^{(\theta^2_0+\theta^2_t  -\theta^2_1 -\theta^2_{\infty})/2} \cdot  \,
(1-t)^{(\theta^2_t+\theta^2_1-\theta^2_0  -\theta^2_{\infty})/2} \cdot \, \tau(t)\right)
\nonumber\\
&&\hspace{-.3in}
  = \, \, t \cdot  \, (t-1) \cdot \, \frac{d}{dt}\ln\left(t^{(n_1n_2+n_3n_4)/2}
\cdot \, (1-t)^{(n_1n_2-n_3n_4)/2}\tau(t) \right), 
\end{eqnarray}
with
\begin{eqnarray}
\label{tau}
&&\hspace{-.3in}
\tau_N^{(p,p',\eta,\xi)}(t)
\, = \, \, \, (1 \, -t)^{-N \cdot \, (N+p+p')/2} \cdot  \, D_N^{(p,p',\eta,\xi)}
\nonumber\\
\label{tau2}
&&\hspace{.1in}
= \, \, (1-t)^{(n_1+n_2)(n_3-n_4)/2} \cdot \, D_N^{(\nu,\nu',\eta,\xi)}, 
\end{eqnarray} 
satisfies the Okamoto equation (\ref{okacanonical}) with
\begin{equation}
\label{okatheta}
\hspace{.16in}
n_1\,\,  =\,\,  \, \theta_t -\theta_{\infty}, \quad \, 
n_2 \,\,  = \,\,  \,\theta_t +\theta_{\infty}, \quad \, 
n_3\, \, =\,\,  \, \theta_0 -\theta_1, \quad \, 
n_4 \,\,  = \, \,\,   \theta_0 +\theta_1, 
\end{equation}
where
\begin{equation}
\label{thetaparam}
\hspace{-.55in} \quad 
(\theta_0, \, \, \theta_t, \, \, \theta_1, \, \, \theta_{\infty}) \, \, = \, \, \,
 \frac{1}{2} \cdot  \, (\eta, \, N, \, -N \, -p-p', \,  \,p \, \, -p'+\eta), 
\end{equation}
are the eigenvalues of the linear system for isomonodromic deformation
\begin{equation}
\label{dY}
\hspace{-.7in} 
\frac{dY}{dz}\,\, = \, \,\,
  \left(\frac{A_0}{z} \, +\frac{A_t}{z-t} \, +\frac{A_1}{z-1}\right) \, Y, 
\end{equation}
where $\, A_i$ are traceless $\, 2\, \times \,  2 \, $ matrices and:
\begin{equation}
\label{AAAA}
\hspace{-.7in} 
A_{\infty} \, \, = \, \,  \,-A_0 \,  \,-A_t \,  \,-A_1. 
\end{equation}
Using (\ref{thetaparam}) in (\ref{okatheta}) we have:
\begin{eqnarray}
&&\hspace{-.5in} \, \, \,
n_1 \,\, = \, \, \, (N \, -p+p'-\eta)/2, \quad \quad \quad
n_2 \,\, = \, \,\,  (N \, +p-p'+\eta)/2,
\nonumber\\
\label{nparam}
&&\hspace{-.5in} \, \, \,
n_3 \,\, =\, \, \, (\eta +N +p+p')/2, \quad\quad
\quad \, n_4\,\, = \, \,\,   (\eta -N -p-p')/2.
\end{eqnarray}
Thus we see, for $\, M+N$  {\em even}, that the parameters $\, n_k$ of
(\ref{okam}) agree with the parameters $\, n_k$ (\ref{nparam})  if
\begin{equation}
\label{evenvvp}
\hspace{-.4in}
\eta \, =\, \, 0, \quad \quad p \, =\, \, \frac{M-N \, +1}{2},
 \quad\quad  \, p' \, = \,\,  \frac{M-N \, -1}{2} 
\end{equation}
and for $\, M+N$  {\em odd} the parameters $\, n_k$ of (\ref{okam}) agree
with the parameters $\, n_k$ (\ref{nparam}) if:
\begin{equation}
\label{oddvvp}
\hspace{-.8in}
\eta\,\, = \,\,  0, \quad \quad \quad \quad
p \, = \, \, p' \, = \, \,\, \frac{M-N}{2}.
\end{equation}
For either choice we see that (\ref{tau2}) reduces to:
\begin{equation}
\label{taudn}
\hspace{-.98in}
\tau_N \, \,\, = \,\,  \, \,  (1-t)^{-MN/2} \cdot \, D_N.
\end{equation}

\vskip .1cm

\subsection{The case $\,M = \,0, \, \, \,N = \,1$}
\label{subthecase}

When $\,M= \,0 \, $ and $\,N=\,1 \, $ we see from (\ref{a2})
with $\, \xi =\,0\, $ that:
\begin{equation}
\label{D1}
\hspace{-.98in}
\quad 
D_1\, = \, \, \, A_0(t)\, \, = \,\,\, 
{}_2F_1\Bigl([\frac{1}{2},\frac{1}{2}], \, [1], \, \,t\Bigr).
\end{equation}
Thus from (\ref{specialc01m}) and (\ref{taudn})
\begin{equation}
\label{c01d1}
\hspace{-.8in}
\quad 
C(0,1)\,\,  = \,\, \, \,  (1\, -t)^{1/2} \cdot \,D_1
  \,\,  =\,\,\,  (1\, -t)^{1/2} \cdot \,\tau_1, 
\end{equation}
and (\ref{htau}) reduces to
\begin{eqnarray}
&&\hspace{-.5in} \quad 
h\,\, = \,\,\,
t \cdot \,(t\, -1) \cdot \,
\frac{d}{dt} \ln \left(t^{1/8} \cdot  \,
(1  \, -t)^{1/8}  \cdot \, (1\, -t)^{-1/2} \cdot \, C(0,1)\right)
\nonumber\\
&&\hspace{-.5in} \quad \quad 
\,\, = \,\,\, t \cdot  \, (t-1) \cdot \, \frac{d}{dt}\ln C(0,1)\, \,\, \, \,
    -\frac{t}{4} \, \,\,\,  \, -\frac{1}{8}, 
\end{eqnarray} 
which agrees with (\ref{shiftcmnm}) as required.

\vskip .1cm

\subsection{The general case}
\label{subgeneral}
  
For  $\, C(M,N)\, $ for $\, \nu \, = \, -k\, $ the special case
(\ref{c01d1}) generalizes to:
\begin{equation}
\label{cmndn}
\hspace{-.6in}
C(M,N) \,\,  =\,\, \, (1-t)^{[(N-M)^2 +1-(1+(-1)^{M+N})/2]/4} \cdot \, D_N.  
\end{equation}
To verify (\ref{cmndn}) we use (\ref{taudn}) to write:
\begin{eqnarray}
&&\hspace{-.4in}
C(M,N)\,\,  =\, \, \, (1-t)^{[N^2+M^2+1-(1+(-1)^{M+N})/2]/4} \cdot \, \tau_N
\nonumber \\
&&\hspace{-.1in} \quad  \quad 
\,\, =\,\, \, (1-t)^{n_1n_2-n_3n_4+1/4} \cdot \, \tau_N.  
\end{eqnarray}
Thus, substituting into (\ref{htau}) we find
\begin{eqnarray}
&&\hspace{-.98in}\,\,\,\,\,\,
h\, = \,\, t \cdot \,(t-1) \cdot \,
\frac{d}{dt}\ln\left(C(M,N) \, t^{(n_1n_2+n_3n_4)/2} \, (1-t)^{-(n_1n_2-n_3n_4)/2-1/4}\right)
\nonumber\\
&&\hspace{-.68in} 
\,\,=\,\, t \cdot \, (t-1) \cdot \, \frac{d}{dt}\ln C(M,N)
 \, + (n_3n_4 \, -\frac{1}{4}) \cdot \, t \, \,
-\frac{1}{2} \cdot \, (n_1n_2 +n_3n_4)
\\
&&\hspace{-.98in}
\,\, \,\,\,\, =\,\, \, t  \cdot \, (t-1) \cdot \, \frac{d}{dt}\ln C(M,N)\,\,\,
-\frac{M^2 +1}{4} \cdot \, t
\,-\frac{1}{8}\cdot \, \Bigl(N^2-M^2\, -{{1 \, +(-1)^{M+N}} \over {2}} \Bigr),
 \nonumber
\end{eqnarray}
which agrees with (\ref{shiftcmnm}) as required.

Finally it may be verified that (\ref{cmndn}) satisfies the boundary
condition (\ref{cmnl}). 

\vskip .1cm

\subsection{Specialization of $ \, D_N$}
\label{subSpec}

It remains to specialize the matrix elements $\,A_m$ of (\ref{a2}) to
the special cases (\ref{evenvvp}) and (\ref{oddvvp}).
For $\, m \,\leq \, 0\, $ we may directly set $\, \eta\,= \, 0$ in (\ref{a2})
and use the identity
\begin{equation}
\label{gammaident}
\hspace{-.8in}
\frac{\Gamma(1+p') \cdot \,\Gamma(-p')}{\Gamma(1+m+p') \cdot  \,\Gamma(-p'-m)}
\,\,\, =\,\,\, (-1)^m, 
\end{equation}
to find:
\begin{equation}
\label{ammm}
A_{-]m]}\,\, = \, \,\,\,
    \frac{\Gamma(|m|-p') \cdot \, (-1)^{|m|} \cdot \, t^{|m|/2}}{|m|!
      \cdot \, \Gamma(-p')} \cdot \, 
{}_2F_1([-p, -p'+|m|], \, [1+|m|], \, \, t).
\end{equation}
For $\, m \,\geq\, 1 \, $ the factor $\,\Gamma(1+\eta-m)\, $ in the denominator
diverges for $\, \eta \, \rightarrow\, 0 \, $ and consequently the
limit $\, \eta \, \rightarrow\,  0\,\, $ must be taken carefully. Then,
using the identity (\ref{gammaident}),
we find for $\, m\, \geq\,  1\, $ as $\, \eta\, \rightarrow \, 0$:
\begin{equation}
\label{ammp}
\hspace{-.55in}
A_m\,\,  = \,\,\,
\frac{\Gamma(-p+m) \cdot \, (-1)^m \cdot \, t^{m/2}}{\Gamma(-p) \cdot \, m!} \cdot \,
{}_2F_1([m-p, -p'], \, [m+1], \,\,t).
\end{equation}
For $\,M+N$ even we thus use (\ref{evenvvp}) in (\ref{ammm}) 
for $\, m\,\leq\, 0 \, $ to find
\begin{eqnarray}
\label{aemm}
&&\hspace{-.6in} \quad 
A_{-|m|}\,\, = \, \, \,
\frac{\Gamma(|m|+\frac{N-M+1}{2}) \cdot \, (-1)^m \cdot \, t^{|m|/2}}
{\Gamma(\frac{N-M+1}{2}) \cdot \, |m|!}
\nonumber \\ 
&&\hspace{-.5in} \quad \quad \quad 
\times \,
{}_2F_1\Bigl([\frac{N-M-1}{2},\frac{N-M+1}{2}+|m|], \, [1+|m|], \, \, t\Bigr), 
\end{eqnarray}
and for $\, m\, \geq \, 1\, $ we use (\ref{evenvvp}) in (\ref{ammp}) to find:
\begin{eqnarray}
\label{aemp}
&&\hspace{-.4in}
A_m \, \, = \,\, \, \,
\frac{\Gamma(m+\frac{N-M-1}{2}) \cdot \, (-1)^m \cdot \, t^{m/2}} 
{\Gamma(\frac{N-M-1}{2}) \cdot \, m!} 
\nonumber \\ 
&&\hspace{-.5in} \quad \quad \quad 
\times \,
{}_2F_1\Bigl([\frac{N-M+1}{2},\frac{N-M-1}{2}+m], \, [1+m], \, \, t\Bigr).
\end{eqnarray}
For $\, M+N$ {\em odd} we use (\ref{oddvvp}) in (\ref{ammm}) and
(\ref{ammp}) to obtain:
\begin{eqnarray}
\label{nao}
\hspace{-1in}
&&\hspace{-.4in}
A_{-m}\, = \, \, A_m \,  \,= \, \, \,
\frac{\Gamma(m+\frac{N-M}{2}) \cdot \, (-1)^m \cdot \, t^{m/2}}  
     {\Gamma(\frac{N-M}{2}) \cdot \, |m|!}
\nonumber     \\
&&\hspace{-.4in} \quad \quad  \quad 
     \times \, 
{}_2F_1([\frac{N-M}{2},\frac{N-M}{2}+m], \, [1+m],  \, \, t).
\end{eqnarray}
Similar Toeplitz elements
can be found for the correlation
functions obtained for $\, T > \, T_c \, $
and $\, \nu=\, -k_>$.
The expressions of $A_m$ for this case
are given in~\ref{ToeplitzTsupTc}.

We note that from (4.35) of~\cite{gil} that for $\, \eta=\, 0\, $ that the
matrix elements are obtained from:
\begin{eqnarray}
\label{etazeroels}
&&\hspace{-.2in}
A_m\, \, =\, \, \, \frac{1}{2\pi}\, \int_0^{2\pi} \, d\theta \, \, 
e^{im\theta} \cdot \, (1 \, -ke^{i\theta})^p \, (1 \, -ke^{-i\theta})^{p'}.
\end{eqnarray}

\vskip .1cm

\subsection{Direct proof for $\, C(0,2)$ when $\, \nu \, = \, -k$}
\label{Direct}

The relation of $\,C(M, N)$ for $\, \nu \, = \, -k$ to the determinant
$\,D_N$ (\ref{cmndn}) was obtained from the nonlinear equations for $\, C(M,N)$
for $\, \nu \, = \, -k$ and $\,D_N$. In this
section we give a direct proof of (\ref{cmndn}) for $\,C(0,2)$, when
$\, \nu \, = \, -k$,  by use of contiguous relations for hypergeometric
functions. This provides a proof of the Okamoto equation (\ref{okacanonical})
with (\ref{okam}) and (\ref{shiftcmnm})  for $\,C(0,2)$  for $\, \nu \, = \, -k$
for $\,T\,<\,T_c$.  The relation
for $\,C(0,1)$    has already been shown in section \ref{subthecase}.

To prove (\ref{cmndn}) for $\, C(0,2)$ we need to prove the following
identity between the $\,2 \,\times \, 2$ determinants for $\, C(0,2)$ and $\,D_2$ 
obtained by using (\ref{aeven}) and (\ref{aodd}) for $\, C(0,2)$ and
(\ref{aemm}) and (\ref{aemp}) for $\, D_2$ 
\begin{eqnarray}
&&\hspace{-.7in}
\begin{array}{|ll|}
\sqrt{1 \, -t} \cdot \,
_2F_1([\frac{1}{2},\frac{1}{2}], \, [1],\,\, t)& -\frac{t^{1/2}}{2}
\cdot \, _2F_1([\frac{1}{2},\frac{1}{2}], \, [2], \,\,t)
\\
\frac{t^{1/2}}{2} \cdot \, _2F_1([\frac{1}{2},\frac{1}{2}], \, [2], \,\,t)&\sqrt{1 \, -t}
  \cdot \, _2F_1([\frac{1}{2},\frac{1}{2}], \, [1], \,\,t)
\end{array}
\nonumber\\
&&\hspace{-.6in}
\,\, = \,\, \, (1-t) \cdot \, 
\begin{array}{|ll|}
  _2F_1([\frac{1}{2}.\frac{3}{2}], \, [1], \,\,t)& -\frac{t^{1/2}}{2}
  \cdot \, _2F_1([\frac{3}{2},\frac{3}{2}], \, [2], \,\, t)
\\
-\frac{3\, t^{1/2}}{2} \cdot \,
_2F_1([\frac{1}{2},\frac{5}{2}], \, [2], \,\,t)& _2F_1([\frac{1}{2},\frac{3}{2}], [1], \,\, t)
\end{array}
\, , 
\end{eqnarray}
which we rewrite 
\begin{eqnarray}
 &&\hspace{-.5in}
  \begin{array}{|ll|}
_2F_1([\frac{1}{2},\frac{1}{2}], \, [1], \,\,t)&\frac{t}{2}
  \cdot \,  _2F_1([\frac{1}{2},\frac{1}{2}], \, [2], \,\,t)
\\
-\frac{1}{2} \cdot \, _2F_1([\frac{1}{2},\, \frac{1 }{2}], \, [2], \,\, t)&(1-t)
\cdot \, _2F_1([\frac{1}{2},\frac{1}{2}], \, [1], \,\, t)
\end{array}
\nonumber\\
&&\hspace{-.3in}
= \, \,  \, 
\begin{array}{|ll|}
(1-t) \cdot \, _2F_1([\frac{1}{2}.\frac{3}{2}], \, [1],  \, \, t)
 &\frac{t\,(1-t)}{2} \cdot \,  _2F_1([\frac{3}{2},\frac{3}{2}], \, [2], \, \, t)
\\
\frac{3}{2} \cdot \, _2F_1([\frac{1}{2},\frac{5}{2}], \, [2], \, \, t)&
_2F_1([\frac{1}{2},\frac{3}{2}], \, [1], \,\, t)
\end{array}
\, . 
\label{proof1}
\end{eqnarray}
We then use (41) on page 103 of~\cite{batemanv1}
\begin{eqnarray}
\label{ident1}
&& \hspace{-.5in}
(1-t) \cdot \, _2F_1\Bigl([\frac{1}{2},\frac{3}{2}], \, [1], \, \, t\Bigr)
\nonumber \\
&& \hspace{-.5in} \quad \quad \quad 
=\,\,\,(1-t) \cdot \, _2F_1\Bigl([\frac{1}{2},\frac{1}{2}], \, [1], \, \, t\Bigr)
\,\,\,
+\frac{t}{2} \cdot \, _2F_1\Bigl([\frac{1}{2},\frac{1}{2}], \, [2],  \, \, t\Bigr), 
\end{eqnarray}
to rewrite the $(1,1)$ element of the righthandside of (\ref{proof1}) and
we use (33) on page 103 of \cite{batemanv1} with
$\,a =\, 1/2, \,b=\, 3/2, \, c=\,2$
\begin{equation}
\hspace{-.8in}
  (1-t) \cdot \, _2F_1\Bigl([\frac{3}{2},\frac{3}{2}], \, [2], \,\, t\Bigr)
\, \, = \,\, \,
_2F_1\Bigl([\frac{1}{2},\frac{1}{2}], \, [2], \, \, t\Bigr), 
\end{equation}
to rewrite the  $(1,2)$ element on the righthandside. Then we subtract column 2
from column 1 to find that the righthand side of (\ref{proof1})
becomes:
\begin{eqnarray}
&& \hspace{-.8in}
  \begin{array}{|ll|}
(1\, -t) \cdot \, _2F_1([\frac{1}{2},\frac{1}{2}], \, [1], \,\, t)&\frac{t}{2}
 \cdot \, _2F_1([\frac{1}{2},\frac{1}{2}], \, [2], \,\,t)
\\
\frac{3}{2}\cdot \, _2F_1([\frac{1}{2},\frac{5}{2}], \, [2],\,\, t)
\, \, - \,  _2F_1([\frac{1}{2},\frac{3}{2}], \, [1], \, \,t)&
_2F_1([\frac{1}{2},\frac{3}{2}], \, [1], \, \, t)
\end{array}
\\
&& \hspace{-.8in} = \, \, \,
\begin{array}{|ll|}
_2F_1([\frac{1}{2},\frac{1}{2}], \, [1],  \, \,t)&\frac{t}{2}
\cdot \, _2F_1([\frac{1}{2},\frac{1}{2}], \, [2], \, \,t)
\\
\frac{3}{2} \cdot \, _2F_1([\frac{1}{2},\frac{5}{2}], [2], \, \,t)
\, - \,  _2F_1([\frac{1}{2},\frac{3}{2}], \, [1], \, \,t)&
(1-t) \cdot \, _2F_1([\frac{1}{2},\frac{3}{2}], \, [1], \, \,t)
\end{array}  \, .
\nonumber
\label{proof2}
\end{eqnarray}
We then rewrite the  $(2,2)$ element on the right hand side (\ref{proof2}) 
using (\ref{ident1}) and subtract row 1 from row 2 to obtain:
\begin{equation}
\begin{array}{|ll|}
  _2F_1([\frac{1}{2},\frac{1}{2}], \, [1], \, \, t)&\frac{t}{2}
  \cdot \, _2F_1([\frac{1}{2},\frac{1}{2}], \, [2], \, \, t)
\\
\frac{3}{2} \cdot  \, _2F_1([\frac{1}{2},\frac{5}{2}], \, [2], \,\, t)
\, \, -  \, _2F_1([\frac{1}{2},\frac{3}{2}], [1],\,\, t)
\,-\,  _2F_1([\frac{1}{2},\frac{1}{2}], [1],\,\, t)& (1-t) \cdot \,
  _2F_1([\frac{1}{2},\frac{1}{2}], \, [1], \,\,t)
\label{proof3}
\end{array} \, .
\nonumber 
\end{equation}
Then  we note that if 
\begin{eqnarray}
\label{proof4}
&& \hspace{-.4in}
\frac{3}{2}\cdot \, _2F_1\Bigl([\frac{1}{2},\frac{5}{2}], \, [2], \, \,t\Bigr)
\, \,\,\, - \, _2F_1\Bigl([\frac{1}{2},\frac{3}{2}], [1], \, \,t\Bigr)\,\,\,
- \, _2F_1\Big([\frac{1}{2},\frac{1}{2}], \, [1], \,\, t\Bigr)
\nonumber \\
&& \hspace{-.4in} \quad \quad  \quad 
\,=\,\, \,
-\frac{1}{2}\cdot  \, _2F_1\Bigl([\frac{1}{2},\frac{1}{2}], \, [2], \, \,t\Bigr), 
\end{eqnarray}
then (\ref{proof3}) agrees with the left hand side of
(\ref{proof1}) as required. To prove (\ref{proof4}) we use (43) on page
104 of~\cite{batemanv1} (with a missing factor of $\, z$ restored in the
  last term) with $\,\, a =\, 1/2,\, \,b= \, 5/2,\,\, \, c = \, 1$
  \begin{equation}
 \hspace{-.02in}
  \frac{1-t}{2}  \cdot \, _2F_1\Bigl([\frac{1}{2},\frac{5}{2}], \, [1], \, \, t\Bigr)
  \, - \, _2F_1\Bigl([\frac{1}{2}, \frac{3}{2}], \, [1], \, \, t\Bigr) \, \, 
  +\frac{t}{2} \cdot  \, _2F_1\Bigl([\frac{1}{2},\frac{5}{2}], \, [2],  \, \, t\Bigr)
  \, = \, \, 0, 
\end{equation}
and (41) on page 103 with $ \,\,a = \,1/2, \,\,b= \,1/2, \, \,c= \,1$
\begin{equation}
\hspace{-.02in}
\frac{1-t}{2}  \cdot \, _2F_1\Bigl([\frac{1}{2},\frac{1}{2}], \, [1],  \, t\Bigr) \, 
-\frac{1-t}{2} \cdot \, _2F_1\Bigl([\frac{1}{2},\frac{3}{2}], \, [1], \, t\Bigr)\, 
+\frac{t}{4} \cdot \, _2F_1\Bigl([\frac{1}{2},\frac{1}{2}], \, [2], \,\, t\Bigr)
=\, 0, 
\end{equation}
to eliminate $\, _2F_1([\frac{1}{2}, \frac{5}{2}], \, [2], \,  \, t)$ and 
$\, _2F_1([\frac{1}{2}, \, \frac{1}{2}], \, [2], \, \, t)$.
The desired result is then obtained by use of (29) on page (103)
of~\cite{batemanv1} with $ \, a =\, 1/2,\, \, b =\, 3/2,\, \, c =\, 1$:
\begin{eqnarray}
&& \hspace{-.4in} \quad  \quad 
\frac{1}{2} \cdot \, _2F_1\Bigl([\frac{1}{2},\frac{1}{2}], \, [1], \, \, t\Bigr)
  \,  \, +(2\, -t) \cdot \, _2F_1\Bigl([\frac{1}{2},\frac{3}{2}], \, [1], \,\, t\Bigr)
\nonumber \\
&& \hspace{-.4in} \quad  \quad  \quad  \quad \quad  \quad 
-\frac{3}{2} \cdot \, (1-t) \cdot \,
_2F_1\Big([\frac{1}{2},\frac{5}{2}], \, [1], \, \, t\Bigr)
  \, \,   = \, \,\, 0.
\end{eqnarray} 

\vskip .1cm

\section{Factorizations}
\label{Factoriz}

All symmetric $N\times N$ Toeplitz determinants can be factored into the product
of two determinants by use of the procedure used by Wilf~\cite{wilf}
for determinants with $N$ even of subtracting column $j$ from
column $\, N+1-j$ for $ \, 1 \, \leq \, j \leq \, N/2 \, $
and then adding row $ \, N+1-j$ to row $j$
for $1\leq j \leq N/2$. Thus, for example, we find\footnote[1]{Note that
the factors in $ \, D_2,\, D_3$ and $\, D_4$ can be put into a Toeplitz form.}:
\begin{eqnarray}
D_2\, = \, \,  \, (A_0-A_1) \cdot  \,(A_0+A_1),
  \\
\vspace{.1in}
D_3\, = \,\,  \, (A_0-A_2)\cdot
  \begin{array} {|cc|}
A_0+A_2&2A_1\\
A_1&A_0\\
  \end{array},
 \\
\vspace{.1in}
D_4 \, = \, \,  \, 
\begin{array}{|ll|}
A_0+A_3&A_1+A_2\\
A_1+A_2&A_0+A_1
\end{array}
\cdot
\begin{array}{|ll|}
A_0-A_1&A_1-A_2\\
A_1-A_2&A_0-A_3\\
\end{array}, 
\\
\vspace{.1in}
D_5\, = \, \,  \, 
\begin{array}{|ccc|}
A_0+A_4&A_1+A_3&2A_2\\
A_1+A_3&A_0+A_2&2A_1\\
A_2&A_1&A_0\\
\end{array}
\cdot
\begin{array}{|ll|}
A_0-A_2&A_1-A_3\\
A_1-A_3&A_0-A_4\\
\end{array}, 
\end{eqnarray}
where the $\, D_n$ and $A_n$ are given by (\ref{det}) and (\ref{a1}). 
Thus for the special case (\ref{nao}) when $\, T <\, T_c$ with $\, M+N$ {\em odd}
$\, M\, < \, N$ and $\, M\, \ne \,  0$, the correlations $\, C(M,N)$ factor
into two factors. 

For the special case (\ref{nao}) when $\, T <\, T_c$ with $\, M+N$ {\em odd}, we find
when $\, M\, = \, 0$, that $\, D_3$ further
factors into three factors so that
\begin{equation}
\hspace{-.53in} \quad \quad 
C(0,3)\,\, =\, \,\,
-4 \cdot \, (1-t)^{1/2} \cdot \, t^{-2} \cdot \, {\tilde E}\cdot({\tilde E}
  \, \, -{\tilde K}) \cdot ({\tilde E}\, \, -(1\, -t) \cdot \,{\tilde K}),   
\end{equation}
and $\, D_N$ with $\, N$ odd, $\, N \, \geq\,  5$,  factors into {\em four} factors, 
so that
\begin{equation}
\hspace{-.5in}
 C(0,N)\, \, = \, \,\, 
 {\rm constant} \cdot \, (1\, -t)^{1/2}
 \cdot \,  t^{(1-N^2)/4} \cdot \, f_1 \, f_2 \, f_3 \, f_4, 
\end{equation}
For example, for $\, C(0,5) $, the $\, f_i$'s read:
\begin{eqnarray}
&&\hspace{-.98in} \quad \,\, \,  
 f_1\,\, =\,\, \, (2t\, -1) \cdot \, {\tilde E}\,\, +(1\, -t) \cdot \, {\tilde K},
 \quad  \,\, \,\,\,\,
f_2\,\, =\,\,\,  (1\, +t) \cdot \, {\tilde E}\,\, -(1\, -t) \cdot \, {\tilde K},
\\
&&\hspace{-.98in} \quad \quad \quad  \quad \quad 
f_3\, =\,\, \,  (t\, -2) \cdot \, {\tilde E}\,\, 
+2 \cdot \, (1\, -t) \cdot \, {\tilde K},
   \, 
  \\
&&\hspace{-.98in} \quad \quad \quad  \quad \quad
f_4\, =\,\,\,  3 \, {\tilde E}^2\,\, +2 \cdot \, (t\, -2) \cdot \,{\tilde E}\, {\tilde K}
\,\,  +(1-t) \cdot \, {\tilde K}^2. 
\end{eqnarray}
We have studied these four factors of $\, C(0,N)$ with $\, N$ odd by the process 
previously described and found that all four factors satisfy the equation
(\ref{okacanonical}) with the identical Okamoto parameters
\begin{equation}
 \hspace{-.6in} \, \,\,  \, 
 n_1\,\, =\, \,\,  \frac{N-1}{4}, \quad \,  \,\,\,\, n_2\,\, =\,\,\, \frac{N+1}{4},
 \quad\,\,\,\, \,  n_3\,\, =\,\, \, -\frac{1}{2}, \quad \,  \,\,\,\, n_4\,\, =\,\, \,0, 
\end{equation}
where the relation of the factors $\, f_j$ to $\, h$ is given by:
\begin{eqnarray}
\label{h1}
&&\hspace{-.4in}
h_1\, \, =\, \,\,  t\cdot   \, (t\, -1) \cdot \, \frac{d\ln f_1}{dt}
\, \,\,\,  -\frac{N^2\,+3}{16} \cdot \, t\, \,\, +\frac{N^2\, +3}{32}, 
\\
\label{h2}
&&\hspace{-.4in}
h_2\, \, =\, \,\,  t  \cdot \, (t\, -1) \cdot \,\frac{d\ln f_2}{dt}
\, \,\, \,-\frac{N^2\,-1}{16} \cdot \, t\, \, \,+\frac{N^2\,+3}{32}, 
\\
\label{h3}
&&\hspace{-.4in}
h_3\, \, =\, \,\,  t  \cdot \, (t\, -1)\cdot \,\frac{d\ln f_3}{dt}
\, \,\,\,  -\frac{N^2\,-1}{16} \cdot \, t\, \,\, +\frac{N^2\,-5}{32}, 
\\
\label{h4}
&&\hspace{-.4in}
h_4\, \, =\, \, \, t \cdot  \, (t\, -1)\cdot \,\frac{d\ln f_4}{dt}
\, \, \,-\frac{N^2\,-5}{16} \cdot \, t\, \,\, +\frac{N^2\,-5}{32}. 
\end{eqnarray}
By comparing (\ref{h1})-({\ref{h4}) for $\, C(0,5)$ with the four cases of boundary
conditions in \ref{boundaryd},  we see that the factors $\, f_1$ and $\, f_2$
are in case 1 with $\, c^{(1)}_0 \, $ and $\, c^{(1)}_1 \, $ given by (\ref{a01}) and
 (\ref{a11}) and from (\ref{cnp11}) the coefficient of $\, t^{(N+3)/2}\, $
is a constant which must be specified separately for $\, f_1$ and
$\, f_2$. Similarly the factors $\, f_3$ and $\, f_4$ are in case 4 with
$\, c^{(4)}_0\, $ and $\, c^{(4)}_1\, $ given by (\ref{a04}) and (\ref{a14}) and
from (\ref{cnp14}) the coefficient of $\, t^{{N+1}/2}\, $ is a constant
which must be specified separately for $\, f_3$ and $\, f_4$.

\vskip .1cm

\section{Discussion}
\label{Discussion}
In this paper we have discovered for the special case $\, \nu=\, -k \, $
and for arbitrary positive integers  $\, M \, \leq \, N$, that the correlation 
$\, C(M,N)$ satisfies an
Okamoto sigma form of the  Painlev{\'e} VI equation (\ref{okacanonical}) with
parameters (\ref{okam}) for $\, T<\, T_c$ and (\ref{okap}), and (\ref{okapodd})
for $\, T >\, T_c$. These non-linear differential equations have been obtained
using extensively Pantone's program
and checked with a large set of exact expressions of the $\, C(M,N)$ in terms
of $\, {\tilde E}$ and  $\, {\tilde K}$. 
Moreover the nonlinear differential equation for $\, T< \, T_c \, $ is 
the same equation satisfied by a particular case of the
$\, N \times\, N\, $ Toeplitz
determinants of Forrester-Witte~\cite{fw} and Gamayun, Igorov and
Lisovyy \cite{gil}. This is perhaps surprising because no Toeplitz form
for $\, C(M,N)$ is in the literature except for $\, M=\, 0$, $\, M=\, N$ and the
results of Au-Yang and Perk~\cite{ayp} for $\, C(N-1,N)$. We have also
investigated in \ref{boundaryd} the boundary conditions
which must be applied to the nonlinear differential equations to obtain solutions
which are analytic at $\, k= \, 0$.

In the course of this investigation we have found several open questions:

\vspace{.1in}

1) We have seen that all the correlations $\, C(M,N)$ considered are
members of a one parameter family of Painlev\'e VI functions but the principle
for determining the specific value of the boundary condition is not known.

\vspace{.1in}

2) In section \ref{Factoriz} we exhibited, for the row correlation
functions, a remarkable phenomenon of a 
Painlev\'e VI sigma function
which satisfies an equation with one set of Okamoto parameters and a
specific boundary condition constant is equal to a {\em sum of four}
\footnote[2]{A similar phenomenon for Painlev{\'e} V with a sigma function
being the sum of {\em two} (see equation (6.23) in~\cite{tracy})
sigma functions was found by Tracy and Widom~\cite{tracy}.} Painlev\'e VI
sigma functions which all have  the same Okamoto parameters (which are 
different from the previous set) and have four specific boundary
condition constants. One would like
to find the conditions yielding such a remarkable phenomenon.

\vspace{.1in}

3) We  also saw in section \ref{Factoriz}  for $\, T \, <\, T_c$, $\, M+N$ odd
and $\, M \, \leq \, N$, 
that $\, C(M,N)$ with $\, M\, \ne \, 0$,
always factors into {\em two} terms. It is not known if
these terms have the Painlev{\'e} property of having no movable
critical points and, if they do have this property,  are they expressible
as known Painlev{\'e} functions ? These factorizable cases need much
further study.

\vspace{.1in}

4) The case $\, M \geq \,  N \, $ with $\, \,  \nu =\, -k\, $ remains
to be understood. By
(\ref{mnsym}) this is equivalent to $\, M \, \leq \,  N$ with $\nu=\, -1/k$
and, when this constraint holds, we see from (\ref{Pidef}) that
$\, {\tilde \Pi}(-\nu \,k,\, k) \, $ becomes singular. This has the effect that
some (but not all) 
correlations for $ \, M\, \geq\,  N$ are no longer homogeneous polynomials 
in $\, {\tilde K}$ and $\, {\tilde E}$. 
For example for $\, T < \, T_c$
\begin{eqnarray}
&&\hspace{-.5in}
  C(2,0) \, \, = \, \, \,\,
  1\, -t \, \,\, \,\, +(1-t) \cdot \, {\tilde K}^2 \, \, 
  \, \, -2 \cdot \, (1-t) \cdot \, {\tilde E}{\tilde K}\, \, 
  +{\tilde E}^2,
  \\
&&\hspace{-.5in}
C(3,0) \, \, =\, \, \,
{\sqrt{1\, -t}} \cdot \, \Bigl[ (1\, -t)^2
\nonumber \\
&&\hspace{-.2in}\, \,  \,
+2 \cdot \, (t\, -1)^2 \cdot \,{\tilde K}^2\, \,
+4 \cdot \, (t\, -1) \cdot \, {\tilde E} {\tilde K}
\, \, +2 \cdot \, (1\, +t) \cdot \, {\tilde E}^2 \Bigr].
\end{eqnarray}
Consequently the corresponding non-linear ODEs
are much more involved than Okamoto sigma form of Painlev\'e VI equations.
It is not even clear that all these non-linear ODEs can be encapsulated
in closed formulae depending on $\, M$ and $\, N$, like
this was the case with the two-parameters
families of equations (\ref{eqnm}), (\ref{EqMNhigh}) and (\ref{eqnp}).

\vspace{.1in}

More generally, the discovery of these four two-parameters families of Okamoto sigma form of
Painlev{\'e} VI equations is a strong incentive to find  non-linear ODEs with the Painlev\'e property,
for two-point correlation functions $\, C(M,N)$ that are not restricted to selected conditions
like $\, \nu \, = \, -k \, $ or $\, \nu \, = \, -1/k \, $ for the anisotropic model.

\vskip .2cm 
\vskip .3cm

{\bf Acknowledgments.}
S. Boukraa would like to thank the LPTMC for hospitality. J-M. Maillard
would like to thank the  Department of Mathematics and Statistics
of The University of Melbourne  for hospitality. J-M.M also thanks C. Cosgrove and
I. Dornic for many fruitful discussions on Painlev\'e equations. We thank
Prof. P. Forrester for providing the reference of Wilf~\cite{wilf}  and
 Prof. M. Jimbo for the reference of Kaneko~\cite{kaneko}.

\vspace{.1in}

\appendix

\section{Calculations of the $\, a_n$'s.}
\label{appFirst}

For $\, a_{2n+1}$ of (\ref{anp}) we first use (36) on page 103
of~\cite{batemanv1} to write
\begin{eqnarray}
&&\hspace{-.48in}
  {}_2F_1\Bigl([n+\frac{1}{2},-\frac{1}{2}], \, [n+1], \, \, \alpha^4\Bigr)
  \,\, = \, \, \,
\frac{1}{2}\cdot \Bigl[  {}_2F_1\Bigl([n-\frac{1}{2}, -\frac{1}{2}], \, [n+1], \, \, \alpha^4\Bigr)
\nonumber\\
&&\hspace{-.2in} \quad \quad 
+(1 \, -\alpha^4) \cdot  \,
{}_2F_1\Bigl([n+\frac{1}{2},\frac{1}{2}], [n+1], \, \,\alpha^4\Bigr)\Bigr], 
\end{eqnarray}
and then by use of (5) on page 111 of~\cite{batemanv1} and
\begin{equation}
\label{kalphap}
\hspace{-.48in}
k_>\, = \, \, \frac{2\, \alpha}{1 +\alpha^2}, 
\end{equation}
we find:
\begin{eqnarray}
&&\hspace{-.98in}
{}_2F_1\Bigl([n+\frac{1}{2}, -\frac{1}{2}], \, [n+1], \, \, \alpha^4\Bigr)
\,  \,= \, \, \, \frac{1}{2 \cdot \, (1+\alpha^2)^{2n-1}} \cdot \, 
{}_2F_1\Bigl([n-\frac{1}{2}, n+\frac{1}{2}], \, [2n+1], \, \, k_>^2\Bigr)
\nonumber\\
\label{result}
&&\hspace{-.2in}  \, \, 
+\frac{1 -\alpha^4}{2 \cdot \, (1\,  +\alpha^2)^{2n+1}} \cdot \,
{}_2F_1\Bigl([n+\frac{1}{2},n+\frac{1}{2}], \, [2n+1], \, \, k_>^2\Bigr). 
\end{eqnarray}
Using (\ref{result}) in (\ref{anp}) we obtain:
\begin{eqnarray}
&&\hspace{-.78in} \, \, \, 
  {{a_{2n+1}} \over { \alpha}}
  \,  \, = \, \, \,
  \frac{\Gamma(n+\frac{1}{2})}{2\sqrt{\pi}n!} \cdot \, 
\Bigl[
\left(\frac{\alpha}{1+\alpha^2}\right)^{2n-1}
\cdot \,
{}_2F_1\Bigl([n-\frac{1}{2}, n+\frac{1}{2}], \, [2n+1], \, \, k_>^2\Bigr)
\nonumber\\
&&\hspace{-.5in} \, \, \, 
+ \left(\frac{\alpha}{1 +\alpha^2}\right)^{2n+1} \, (\alpha^{-2} -\alpha^2) \cdot \, 
{}_2F_1\Bigl([n+\frac{1}{2},n+\frac{1}{2}], \,[2n+1], \, \, k_>^2\Bigr)\Bigr]. 
\end{eqnarray}
Thus, by using (\ref{kalphap}) and the inverse
\begin{equation}
\label{inverse}
\hspace{-.78in} 
\alpha^{\pm 2} \,\, = \, \,\, \frac{2\, -k_>^2 \,  \, \mp\, 2\, \sqrt{1 -k_>^2}}{k_>^2}, 
\end{equation}
we obtain the final result:
\begin{eqnarray}
\label{finalanp}
&&\hspace{-.98in}
a_{2n+1} \, = \, \, \frac{\Gamma(n+\frac{1}{2})}{\sqrt{\pi}n!} \cdot \, 
\left(\frac{k_>}{2}\right)^{2(n-1)} \cdot \, 4 \cdot \,(1\, -\sqrt{1\,-k_>^2})
\\
&&\hspace{-.98in}
\times \, \Bigl[ {}_2F_1\Bigl([n-\frac{1}{2},n+\frac{1}{2}], \, [2n+1], \, \, k_>^2\Bigr)
+\sqrt{1-k_>^2} \cdot \,
{}_2F_1\Bigl([n+\frac{1}{2},n+\frac{1}{2}], \, [2n+1], \, \, k_>^2\Bigr)\Bigr].
\nonumber
\end{eqnarray}
For $\, \alpha \, \,  a_{-(2n+1)} \, $  (see (\ref{anm})) we proceed   
in a similar fashion, and use (33) on page 103 of~\cite{batemanv1} to write:
\begin{eqnarray}
\label{F=Falpha}  
&&\hspace{-.6in}
{}_2F_1\Bigl([n+\frac{1}{2}, \frac{1}{2} ], \, [n+2], \, \, \alpha^4\Bigr)
  \, \,   = \, \, \, \Bigl(n+\frac{3}{2}\Bigr) \cdot \,
 {}_2F_1\Bigl([n+\frac{1}{2}, -\frac{1}{2}],\, [n+2], \, \, \alpha^4\Bigr)
\nonumber\\
&&\hspace{-.2in}\quad  \, \, \,
-\Bigl(n+\frac{1}{2}\Bigr) \cdot \, (1 -\alpha^4)  
\cdot  \,
 {}_2F_1\Bigl([n+\frac{3}{2}, \frac{1}{2}], \, [n+2], \,  \, \alpha^4\Bigr).
\end{eqnarray}
Then we use (5) on page 111 of~\cite{batemanv1} to obtain
\begin{eqnarray}
&&\hspace{-.78in} \quad 
  {}_2F_1\Bigl([n+\frac{1}{2},\frac{1}{2}], \, [n+2],  \, \alpha^4\Bigr) \,
\nonumber\\
&&\hspace{-.7in} \quad \quad 
= \, \, \Bigl(n+\frac{3}{2}\Bigr) \cdot \, (1 \, +\alpha^2)^{-2n-1} \cdot \, 
{}_2F_1\Bigl([n+\frac{1}{2}, n+\frac{3}{2}],\, [2n+3], \, \, k_>^2\Bigr)
\nonumber\\
&&\hspace{-.7in} \quad \quad 
-\Bigl(n+\frac{1}{2}\Bigr) \cdot \, \frac{1 -\alpha^4}{(1 +\alpha^2)^{2n+3}} \cdot \, 
{}_2F_1\Bigl([n+\frac{3}{2}, n+\frac{3}{2}], \, [2n+3], \, \, k_>^2\Bigr), 
\end{eqnarray}
and thus
\begin{eqnarray}
  \label{alphaa}
&&\hspace{-1in}
\alpha\, a_{-(2n+1)}
\, \, = \, \,
\\
&&\hspace{-1in}
\, \, \, \, \, 
\alpha^2 \cdot \, \frac{\Gamma(n+\frac{1}{2})}{2\sqrt{\pi}(n+1)!} \cdot \, 
  \Bigl[
\left(\frac{\alpha}{1 +\alpha^2}\right)^{2n+1} \cdot \, 
\Bigl(n+\frac{3}{2}\Bigr) \cdot \,
  {}_2F_1\Bigl([n+\frac{1}{2},n+\frac{3}{2}], \, [2n+1], \, \, k_>^2\Bigr)
\nonumber\\
&&\hspace{-.68in}  
-\left(\frac{\alpha}{1+\alpha^2}\right)^{2n+3}\cdot \, 
\Bigl(n+\frac{1}{2}\Bigr) \cdot \,  (\alpha^{-2} -\alpha^2) \cdot \, 
{}_2F_1\Bigl([n+\frac{3}{2},n+\frac{3}{2}], \, [2n+3], \, \, k_>^2\Bigr)\Bigr],
\nonumber
\end{eqnarray}
which by use of (\ref{kalphap}) and (\ref{inverse}) becomes:
\begin{eqnarray}
  &&\hspace{-.8in}\quad
  \alpha \, a_{-(2n+1)}
 \,\, = \, \, \, \frac{\Gamma(n+\frac{1}{2})}{2 \sqrt{\pi}}
\cdot \, \Bigl({{k_>} \over {2}}\Bigr)^{2n+1} \cdot  \, 
\left(\frac{2 \, -k^2 \, -2 \, \sqrt{1-k_>^2}}{ k_>^2}\right) \, 
\nonumber\\
&&\hspace{-.6in} \, \, \quad 
\times \, 
\Bigl[\Bigl(n+\frac{3}{2}\Bigr) \cdot \,
  {}_2F_1\Bigl([n+\frac{1}{2},n+\frac{3}{2}], \, [2n+3], \, \, k_>^2\Bigr)
\nonumber\\
&&\hspace{-.4in} \quad \quad 
-\Bigl(n+\frac{1}{2}\Bigr) \cdot \, \sqrt{1-k^2} \cdot \,
{}_2F_1\Bigl([n+\frac{3}{2},n+\frac{3}{2}], \, [2n+3], \, \, k_>^2\Bigr)\Bigr]. 
\end{eqnarray}
To obtain the final desired result we first carry out the
multiplication to write
\begin{equation}
\label{aalha}
\hspace{-.6in}
\alpha \cdot \, a_{-(2n+1)} \, \, = \,\,\, \, 
\frac{\Gamma(n+\frac{1}{2})}{8 \, \sqrt{\pi}(n+1)!}
\cdot \, \Bigl({{k_>} \over {2}} \Bigr)^{2n-1} \cdot \, 
\Bigl(T_1\,\,  -\sqrt{1\,-k_>^2} \cdot \,\,T_2\Bigr), 
\end{equation}
with
\begin{eqnarray}
&&\hspace{-.5in}
T_1 \,= \,\,\,
(2\, -k_>^2) \cdot \, \Bigl(n+\frac{3}{2}\Bigr)  \cdot \,
  {}_2F_1\Bigl([n+\frac{1}{2}, n+\frac{3}{2}], \, [2n+3], \,\, k_>^2\Bigr)
\nonumber\\
&&\hspace{-.2in}
+2 \cdot \, (1\, -k^2) \cdot \,  \Bigl(n+\frac{1}{2}\Bigr) \cdot \,
{}_2F_1\Bigl([n+\frac{3}{2},n+\frac{3}{2}], \, [2n+3], \, \, k_>^2\Bigr), 
\end{eqnarray}
and
\begin{eqnarray}
&&\hspace{-.4in}
   T_2\,= \,\, \,
   2 \cdot \, \Bigl(n+\frac{3}{2}\Bigr) \cdot \,
 {}_2F_1 \Bigl([n+\frac{1}{2},n+\frac{3}{2}], [2n+3], \, \,  k_>^2\Bigr)
\nonumber\\
&&\hspace{-.3in} \quad  \quad 
+(2-k_>^2) \cdot \, \Bigl(n+\frac{1}{2}\Bigr) \cdot \, 
{}_2F_1 \Bigl([n+\frac{3}{2},n+\frac{3}{2}],\, [2n+3], \, \,  k_>^2 \Bigr), 
\end{eqnarray}
and then note the identities which may be discovered by use
of series expansions on  Maple
and then proven by the use of contiguous identities
\begin{eqnarray}
&&\hspace{-.3in}
T_1\, =\, \,\,
4 \cdot \, (n+1)\cdot \,
{}_2F_1\Bigl([n-\frac{1}{2},n+\frac{1}{2}],\, [2n+1],\,\, k_>^2\Bigr), 
\\
&&\hspace{-.3in}
T_2\, =\,\,\,
4\cdot \,  (n+1) \cdot \,
{}_2F_1\Bigl([n+\frac{1}{2}, n+\frac{1}{2}], \, [2n+1], \, \, \, k_>^2\Bigr), 
\end{eqnarray}
to find the desired result
\begin{eqnarray}
&&\hspace{-.98in}
a_{-(2n+1)} \,\,  = \, \,\,  \frac{\Gamma(n+\frac{1}{2})}
{\sqrt{\pi}n!} \cdot \, \left(\frac{k_>}{2}\right)^{2(n-1)}
\cdot \, 4  \cdot \, (1 \,+\sqrt{1-k_>^2})
\\
\label{finalanm}
&&\hspace{-.98in}
\times
\Bigl[ {}_2F_1\Bigl([n-\frac{1}{2}, n+\frac{1}{2}], \, [2n+1],  \,  k_>^2\Bigr)
  \,  \,  -\sqrt{1-k_>^2} \cdot \,
{}_2F_1\Bigl([n+\frac{1}{2},n+\frac{1}{2}], \, [2n+1],\,  k_>^2\Bigr)\Bigr],
\nonumber 
\end{eqnarray}
which is to be compared with the result for $ \, a_{2n+1}\, $ of (\ref{finalanp}).

\vskip .1cm

\section{Examples of $\,  C(M,N)$ and ${\tilde C}(M,N)$ for \,  $\nu=\,  -k$}
\label{appA}

For $\,  T <\,  T_c$
\begin{eqnarray}
&&\hspace{-.8in}
C(0,2)\,  \, = \, \, \,
\frac{1}{t} \cdot \, \Bigl[  {\tilde E}^2\,
-2 \cdot \, (1-t) \cdot \,  {\tilde E}{\tilde K}
\,\,  +(1-t) \cdot \,  {\tilde K}^2 \Bigr], 
\\
&&\hspace{-.8in}
C(0,3) \,  \, = \, \,  \,
-\frac{4 \, \, \sqrt{1\, -t}}{t^2} \cdot  {\tilde  E} \cdot ({\tilde E}\, \,
-{\tilde K}) \cdot ({\tilde E} \,\, +(t\, -1) \cdot \, {\tilde K}) 
\\
&&\hspace{-.8in}
C(1,3) \, \,
=\, \, \,
\frac{4}{3\, t^2} \cdot \, \Bigl[  (2 \, -t) \cdot \,{\tilde E}^3\,
-5\cdot \,  (1 \, -t) \cdot \,  {\tilde E}^2 \, {\tilde K}
\nonumber \\
&&\hspace{-.8in}\, \quad  \quad  \quad \quad   \quad
+(1 \, -t) \cdot  \,  (2\, -t) \cdot \,  {\tilde E}{\tilde K}^2
\,  -(1 \, -t)^2 \cdot  \,  {\tilde K}^3 \Bigr] , 
\\
&&\hspace{-.8in}
C(0,5) \,  \, = \, \, \,
\frac{256 \, \, \sqrt{1\, -t}}{81\,  t^6} \cdot
\, \Bigl[(1\, +t) \cdot  \, \tilde{E} \, +(t-1)\cdot  \, \tilde{K}\Bigr]
\cdot \,
\Bigl[(t \, -2)\cdot  \, \tilde{E} \, +2 \, (1 \, -t) \cdot  \, \tilde{K}\Bigr]
 \nonumber\\
 &&\hspace{-.78in}   \, 
 \times \, \Bigl[(2t\,-1) \cdot  \, \tilde{E} \,  \, +(1 \, -t)\cdot  \, \tilde{K}\Bigr]
 \cdot \, \Bigl[ 3 \, \tilde{E}^2 \, \,  +(2 t \, -4)\cdot  \, \tilde{E} \, \tilde{K}
   \, \, \, +(1\, -t) \cdot  \, \tilde{K}^2 \Bigr].
\end{eqnarray}
For $\, \,  T>\,  T_c $
\begin{eqnarray}
&&\hspace{-.98 in}
  C(1,3)\,\,   =\, \, \,
  -\frac{4}{3\,\, t^{5/2}} \cdot \, \Bigl[ (1\, -2t) \cdot \,  {\tilde E}^3 \,\,
-(1 -t) \cdot \,  (3\, -t)\cdot  \,  {\tilde E}^2\,  {\tilde K}\,
\nonumber\\
\label{c13p}
&&\hspace{-.98in} \quad \quad  \quad    \quad  \quad
+(1 -t)\cdot  \,  (3\, -t)\cdot  \,  {\tilde E}\,  {\tilde K}^2
\, \,  -(1\, -t)^2 \cdot  \, {\tilde K}^3\Bigr], 
  \\
\label{c04p}  
&&\hspace{-.98in}
 C(0,4) \,\,   = \,\, \,
 -\frac{16}{9\,\, t^4} \cdot\,  \Bigl[ (5-5t-t^2) \cdot \, {\tilde E}^4 \,\,
   -8\cdot \,  (2-t) \cdot \,  (1-t) \cdot \,  {\tilde E}^3\,  {\tilde K}
   \, \,   +(1\, -t)^3 \cdot \,  {\tilde K}^4 
\nonumber\\
&&\hspace{-.98in} \quad \quad  \quad  \quad 
+2 \cdot \,  (1-t) \cdot  \, (3 \, -t) \cdot \,
(3-2t) \cdot  \,  {\tilde E}^2\,  {\tilde K}^2 
\, \,\,  -4 \cdot \,  (2\, -t) \cdot \,  (1\, -t)^2 \cdot \,  {\tilde E}\,  {\tilde K}^3
\Bigr]
\nonumber\\
&&\hspace{-.98in}\,  \, \quad \quad  \quad  \quad
= \, \, \, \frac{16}{9\,\, t^4}  \cdot\, C_{+} \cdot  C_{-},
\quad  \quad  \quad  \quad  \quad \quad  \quad \hbox{where:}  
\end{eqnarray}
\begin{eqnarray}
&&\hspace{-.98in} \quad  \quad \quad \quad 
C_{\pm} \, \,  = \, \, \, \, 
\Bigl(2\, -t \,\,\,  \pm \, 3 \cdot \, (1\, -t)^{1/2}\Bigr)  \cdot \,   {\tilde E}^2 \, \,
\, \pm (1\, -t)^{3/2} \cdot \, {\tilde  K}^2
\nonumber\\
&&\hspace{-.98in}\,  \quad  \quad \quad \quad  \quad  \quad  \quad  \quad  \quad  \quad 
   -\Bigl(2 \cdot \, (1\, -t) \, \, \pm \,  2 \cdot \, (2\, -t)
     \cdot \, (1\, -t)^{1/2} \Bigr) \cdot \, {\tilde E}\, {\tilde K}. 
\end{eqnarray}
For $ T >\, T_c \, $ with $\, M+N \, $ odd examples of $\, \tilde{C}(M,N)$
of (\ref{tildec}): 
\begin{eqnarray}
&&\hspace{-.98in}
{\tilde C}(1,2)\, \, = \, \, \,
\frac{1}{t} \cdot \, \Bigl[ (t\,-1) \cdot \, {\tilde K}^2 \, \,
  -2  \cdot \, (t\, -2)\cdot \, {\tilde  K} \, {\tilde E}
  \, \, -3 \cdot \, {\tilde E}^2 \Bigr],
\\
&&\hspace{-.98in}
{\tilde C}(1,4)\, \, = \, \, \,
\frac{16}{9 \,  \, t^4} \cdot \, \Bigl[ 5  \cdot \, (t-1)^3 \cdot \, {\tilde K}^4 
\, \, -12  \cdot  \, (t\, -2) \cdot\, (t-1)^2 \cdot \, {\tilde E}{\tilde K}^3
\, \, -4   \cdot \, (t\, -2)^3 \cdot \, {\tilde E}^3  \,{\tilde K}
\nonumber\\
&&\hspace{-.2in}
+6  \cdot \, (t\, -1) \cdot \, (t^2-7t+7) \cdot \, {\tilde E}^2 \, {\tilde K}^2 \, \, 
-9 \cdot \, (1-t+t^2) \cdot \, {\tilde E}^4\Bigr],
\\
&&\hspace{-.98in}
{\tilde C}(2,3) \, \, =\, \, \,
\frac{4}{9\, \,  t^{5/2}} \cdot \,
\Bigl[ (3t \, -1)  \cdot \, (t\, -1)^2 \cdot \, {\tilde K}^3 \, \, 
-(t\, -1) \cdot \, (6t^2-17t+3) \cdot \, {\tilde E} \, {\tilde K}^2
\nonumber\\
&&\hspace{-.2in}
-(20t^2-31t+3) \cdot \, {\tilde E}^2 \, {\tilde K} \, \, 
+(t^2-16t+1) \cdot \, {\tilde E}^3 \Bigr]. 
\end{eqnarray}

\vskip .1cm

\section{Correlation functions as Toeplitz determinants for $\, T>\, T_c$}
\label{ToeplitzTsupTc}

We have shown in section~\ref{Forrester} that the correlation functions
for $\, T<\, T_c\, $ and $\, \nu=\, -k\, $ are given in terms
of Toeplitz determinants (see eqs. (\ref{cmndn}), (\ref{aemm}-\ref{nao}))
using the results of Forrester-Witte~\cite{fw}.

Using the same method, one can easily generalise
these equations to $\,T>\,T_c\, $ and $\,\nu=\, -k_>$.

One verifies that

\begin{itemize}

\item For $\, T>\, T_c\, $ and $\, M+N$ even

\begin{equation}
\label{cmndnTsupeven}
\hspace{-.6in}
C(M,N) \,\, =\,\, \, (-1)^{(N-M)/2} \cdot (1\, -t)^{(N-M)^2/4} \cdot \,  D_N, 
\end{equation}
the Toeplitz matrix elements are given for $\, m \ge \, 1\, $ by:
\begin{eqnarray}
\label{aempsupeven1}
&&\hspace{-.4in}
A_m \, \, = \,\, \, \,
\frac{\Gamma(\frac{N-M-1}{2}+m)}
{\Gamma(\frac{N-M+1}{2}) \, (m-1)! } \cdot (-1)^{m-1} \cdot \, t^{(m-1)/2}
\nonumber \\
&&\hspace{-.5in} \quad \quad \quad
\times \,
{}_2F_1\Bigl([\frac{N-M-1}{2},\frac{N-M-1}{2}+m], \, [m], \, \, t\Bigr)
\end{eqnarray}
and for $ \, m < \,  1$ by
\begin{eqnarray}
\label{aempsupeven2}
&&\hspace{-.4in}
A_m \, \, = \,\, \, \,
\frac{\Gamma(\frac{N-M+1}{2}-m)}
{\Gamma(\frac{N-M-1}{2}) \, (1-m)! } \cdot \, (-1)^{m-1} \cdot \, t^{(1-m)/2}
\nonumber \\
&&\hspace{-.5in} \quad \quad \quad
\times \,
{}_2F_1\Bigl([\frac{N-M+1}{2},\frac{N-M+1}{2}-m], \, [2-m], \, \, t\Bigr).
\end{eqnarray}

\item For $\, T>\, T_c \, $ and $\, M+N$ odd, we know that
$\, C(M,N)=\, 0$. After removing the vanishing factor
$\, (1\, +\nu/k_>)^{1/2}$, one obtains: 
\begin{equation}
\label{cmndnTsupodd}
\hspace{-.6in}
\tilde{C}(M,N) \,\, =\,\, \,
(-1)^{(N-M+1)/2} \cdot \, (M+N) \cdot \, (1\, -t)^{((N-M)^2-1)/4} \cdot \, D_N
\end{equation}

The Toeplitz matrix elements for $ \, m \ge \,  1$ are
\begin{eqnarray}
\label{aempsupodd1}
&&\hspace{-.4in}
A_m \, \, = \,\, \, \,
\frac{\Gamma(\frac{N-M}{2}-1+m) }
{\Gamma(\frac{N-M}{2}) \, (m-1)! } \cdot (-1)^{m-1} \cdot \,  t^{(m-1)/2}
\nonumber \\
&&\hspace{-.5in} \quad \quad \quad
\times \,
{}_2F_1([\frac{N-M}{2},\frac{N-M}{2}-1+m], \, [m], \, \, t), 
\end{eqnarray}
and for $\, m <\,  1$
\begin{eqnarray}
\label{aempsupodd2}
&&\hspace{-.4in}
A_m \, \, = \,\, \, \,
\frac{\Gamma(\frac{N-M}{2}+1-m) }
{\Gamma(\frac{N-M}{2}) \, (1-m)! } \cdot (-1)^{m-1} \cdot \,  t^{(1-m)/2}
\nonumber \\
&&\hspace{-.5in} \quad \quad \quad
\times \,
{}_2F_1\Bigl([\frac{N-M}{2},\frac{N-M}{2}+1-m], \, [2-m], \, \, t\Bigr).
\end{eqnarray}

\end{itemize}

One verifies that these expressions for the correlation functions
are totally compatible with the ones given in \ref{appA} obtained
by the quadratic difference equations (see section~\ref{quadra}).

\vskip .1cm

\section{Boundary conditions}
\label{boundaryd}

The study of solutions of PVI analytic at $t=0$ was first done be
Kaneko~\cite{kaneko} for generic parameters in the Hamiltonian
formalism where four solutions were found. Here we concentrate on the
non-generic cases which allow one parameter families of solutions.
This is most systematically  done by  determining the allowed
boundary conditions on the canonical equation of Okamoto
(\ref{okacanonical}) which are analytic at $ \, t= \, 0$. 
Thus we set in (\ref{okacanonical})
\begin{equation}
\hspace{-.98in}
h(t) \, \, = \, \,  \, \,  \sum_{n=0} \, c_n \cdot \, t^n.
\end{equation}
From the constant term we find
\begin{eqnarray}
&&\hspace{-1in} \quad \quad \quad \,  \, \,
  n_1^2n_2^2n_3^2  \, +n_1^2n_2^2n_4^2 \,  +n_1^2n_3^2n_4^2  \, +n_2^2n_3^2n_4^2
  \,  \, \,-4 \,c_0  \cdot \, n_1 n_2 n_3 n_4
\nonumber\\
&&\hspace{-1in}  \quad \quad\quad \quad   \,  \, \, \, \,
+ c_1 \cdot \,
(n_1^2n_2^2 +n_1^2n_3^2 +n_1^2n_4^2 +n_2^2n_3^2 +n_2^2n_4^2 +n_3^2n_4^2 \, -2 \, n_1n_2n_3n_4)
\nonumber\\
\label{f0}
&&\hspace{-1in}  \quad \quad \quad  \quad \quad  \, \,
+c_1^2  \cdot \,(n_1^2 +n_2^2 +n_3^2 +n_4^2)
   \,\,  \, -4 \cdot \, c_0 c_1 \cdot \, (c_0 +c_1)
\, \,= \,\, \, 0, 
\end{eqnarray}
and from the $ \, t$ term: 
\begin{eqnarray}
&&\hspace{-.8in}
  -2  \cdot \,  c_2 \cdot  \,
  \Bigl[ 4\, c_0^2 \, \,  +8\, c_0 c_1  \, \,
    -2 \, c_1 \cdot \, (n_1^2+n_2^2+n_3^2+n_4^2)
\nonumber\\
\label{f1}
&&\hspace{-.8in} \quad  \quad 
-n_1^2n_2^2 -n_1^2n_3^2 -n_1^2n_4^2 -n_2^2n_3^2 -n_2^2n_4^2 -n_3^2n_4^2
 \, \,  +2 \,n_1 n_2 n_3 n_4 \Bigr] \, \, = \, \, \, 0.
\end{eqnarray}

We solve (\ref{f1}) for $\,  c_1$ in terms of 
$\, c_0$ and $\, n_k$ by setting the term in brackets to zero
and using this in (\ref{f0}) we find that
the resulting fourth order equation in $\, c_0$ factors into four factors
linear in $\, c_0$ and thus we have the four solutions for $ \, c_0$ of
\begin{eqnarray}
\label{a01}
&&\hspace{-.3in}
c^{(1)}_0 \, \, = \, \,  \,
\Bigl(-n_1n_2-n_3n_4 \, +(n_1+n_2)\, (n_3+n_4)\Bigr)/2,
\\
\label{a02}
&&\hspace{-.3in}
c^{(2)}_0 \,\,  = \,\,  \,
\Bigl( n_1n_2+n_3n_4 \, +(n_1-n_2) \, (n_3-n_4) \Bigr)/2,
\\
\label{a03}
&&\hspace{-.3in}
c^{(3)}_0 \,\,  = \, \,\,
\Bigl( n_1n_2+n_3n_4 \, -(n_1-n_2) \, (n_3-n_4) \Bigr)/2,
\\
\label{a04}
&&\hspace{-.3in}
c^{(4)}_0\,\,  = \, \, \,
\Bigl(-n_1n_2-n_3n_4 \, -(n_1+n_2)\, (n_3+n_4)\Bigr)/2, 
\end{eqnarray}
and thus for $ \, c_1$ the companion values are: 
\begin{eqnarray}
\label{a11}
&&\hspace{-.3in}
c^{(1)}_1 \, \, = \, \, \,
\frac{(n_1+n_2) \cdot \,  n_3 n_4 \, \, -n_1 n_2 \cdot \, (n_3+n_4)} {n_1+n_2-n_3-n_4},
\\
\label{a12}
&&\hspace{-.3in}
c^{(2)}_1 \, \, = \, \, \,
\frac{(n_1-n_2)\cdot \, n_3 n_4 \,\,  -n_1 n_2 \cdot \, (n_3-n_4)}{-n_1+n_2+n_3-n_4}, 
\\
\label{a13}
&&\hspace{-.3in}
c^{(3)}_1\, \, =\, \, \,
\frac{(n_1-n_2) \cdot \, n_3 n_4 \,\,  +n_1 n_2 \cdot \, (n_3-n_4)}{-n_1+n_2-n_3+n_4},
\\
\label{a14}
&&\hspace{-.3in}
c^{(4)}_1\, \,= \,\, \,
\frac{(n_1+n_2)\cdot \, n_3 n_4 \,\,  +n_1 n_2 \cdot \, (n_3+n_4)} {n_1+n_2+n_3+n_4}. 
\end{eqnarray}
Case 4 is invariant under all permutations of the $\, n_k$. Case 1
is obtained from case 4 by changing the signs of $\, n_3$ and $\, n_4$. Case
2 is obtained from case 4 by changing the signs of $\, n_1$ and $\, n_4$.  
Case 3 is obtained from case 4 by changing the signs of $\, n_2$ and
$\, n_4$. These sign changes are symmetries of the equation but not of
the solutions.

The values of $\, c_0$ and $\, c_1$ for each of the four solutions may now be
used to compute the term of order $\, t^2$ in the series expansion 
of (\ref{okacanonical})
 and we find that (\ref{okacanonical}) holds if $\, c_2$
 satisfies a linear equation. Thus we obtain:
\begin{eqnarray}
\label{a21}
&&\hspace{-1.01in}
c_2^{(1)} =  
-\frac{(n_1+n_2)(n_1-n_3)(n_1-n_4)(n_2-n_3)(n_2-n_4)(n_3+n_4)}
 {(n_1+n_2-n_3-n_4)^2 (n_1+n_2-n_3-n_4+1) (n_1+n_2-n_3-n_4-1)},  
\\
\label{a22}
&&\hspace{-1.01in}
c_2^{(2)} =  
-\frac{(n_1-n_2)(n_1-n_3)(n_1+n_4)(n_2+n_3)(n_2-n_4)(n_3-n_4)}
  {(n_1-n_2-n_3+n_4)^2(n_1-n_2-n_3+n_4+1)(n_1-n_2-n_3+n_4-1)}, 
\\
\label{a23}
&&\hspace{-1.01in}
c_2^{(3)} = 
\frac{(n_1-n_2)(n_1+n_3)(n_1-n_4)(n_2-n_3)(n_2+n_4)(n_3-n_4)}
    {(n_1-n_2+n_3-n_4)^2(n_1-n_2+n_3-n_4+1)(n_1-n_2+n_3-n_4-1)},
\\
\label{a24}
&&\hspace{-1.01in}
c_2^{(4)} = 
\frac{(n_1+n_2)(n_1+n_3)(n_1+n_4)(n_2+n_3)(n_2+n_4)(n_3+n_4)}
 {(n_1+n_2+n_3+n_4)^2(n_1+n_2+n_3+n_4+1)(n_1+n_2+n_3+n_4-1)}.
\end{eqnarray}

Continuing the recursive procedure we find from the $ \, t^3$ term in
(\ref{okacanonical})
\begin{eqnarray}
&&\hspace{-.98in}
c_3^{(1)} \,\,  =  \,\,\,  2 \cdot \,
\frac{N_3^{(1)} \cdot \, c_2^{(1)} }
     { (n_1+n_2-n_3-n_4)\,(n_1+n_2-n_3-n_4+2)\,(n_1+n_2-n_3-n_4-2)},
\nonumber
\end{eqnarray}
\begin{eqnarray}
\label{a31} 
 &&\hspace{-.8in}
N_3^{(1)} \, \, = \, \, \,
n_1^2n_2-n_1^2n_3-n_1^2n_4-n_2^2n_3-n_2^2n_4 -n_3^2n_4
\nonumber \\
&&\hspace{-.8in} \quad  \quad  \quad  \quad 
+n_1n_2^2+n_1n_3^2 +n_1n_4^2 +n_2n_3^2 +n_2n_4^2 -n_3n_4^2
\nonumber\\
&&\hspace{-.8in}\quad  \quad  \quad  \quad 
-n_1n_2n_3 -n_1n_2n_4 +n_1n_3n_4 +n_2n_3n_4
\, -n_1 -n_2 +n_3 +n_4,
\end{eqnarray}
\begin{eqnarray}
\label{a32}
&&\hspace{-.98in}
c_3^{(2)} \,  =  \,  \,
- 2 \cdot \,  \frac{N_3^{(2)} \cdot \,c_2^{(2)}  }
{(n_1-n_2-n_3+n_4)(n_1-n_2-n_3+n_4+2)(n_1-n_2-n_3+n_4-2)}, 
\nonumber
\end{eqnarray}
\begin{eqnarray}
&&\hspace{-.8in} \quad 
N_3^{(2)}\,\, =\, \,\,
n_1^2n_2 +n_1^2n_3 -n_1^2n_4 +n_2^2n_3 -n_2^2n_4 -n_3^2n_4
\nonumber\\
&&\hspace{-.8in} \quad  \quad  \quad 
-n_1n_2^2 -n_1n_3^2 -n_1n_4^2 +n_2n_3^2 +n_2n_4^2 +n_3n_4^2
\nonumber\\
&&\hspace{-.8in} \quad  \quad  \quad
-n_1n_2n_3 +n_1n_2n_4 +n_1n_3n_4 -n_2n_3n_4 \,\,
+n_1 -n_2 -n_3 +n_4, 
\end{eqnarray}
\begin{eqnarray}
\label{a33}
&&\hspace{-.98in}
c_3^{(3)} \,   = \,  \,
- 2 \cdot \, \frac{N_3^{(3)} \cdot \, c_2^{(3)}}
{ (n_1-n_2+n_3-n_4) \, (n_1-n_2+n_3-n_4 \, +2) \, (n_1-n_2+n_3-n_4 \, -2)}, 
\nonumber
\end{eqnarray}
\begin{eqnarray}
&&\hspace{-.8in} \quad 
N_3^{(3)}\, =\, \,
n_1^2n_2 -n_1^2n_3 +n_1^2n_4 -n_2^2n_3 +n_2^2n_4 +n_3^2n_4
\nonumber\\
&&\hspace{-.8in} \quad  \quad 
-n_1n_2^2 -n_1n_3^2 -n_1n_4^2 +n_2n_3^2 +n_2n_4^2 -n_3n_4^2
\nonumber\\
&&\hspace{-.8in} \quad  \quad 
+n_1n_2n_3 -n_1n_2n_4 +n_1n_3n_4 -n_2n_3n_4 +n_1 -n_2 +n_3 -n_4, 
\end{eqnarray}
\begin{eqnarray}
\label{a34}
&&\hspace{-.98in}
c_3^{(4)} \,  = \, \,
2 \cdot \,  \frac{N_3^{(4)} \cdot \,  c_2^{(4)}}
{ (n_1+n_2+n_3+n_4) \, (n_1+n_2+n_3+n_4\, +2) \,(n_1+n_2+n_3+n_4 \,-2)}, 
\nonumber\\
&&\hspace{-.8in} \quad 
N_3^{(4)} \,\, =\, \,\,
n_1^2n_2 +n_1^2n_3 +n_1^2n_4 +n_2^2n_3 +n_2^2n_4 +n_3^2n_4
\nonumber\\
&&\hspace{-.8in} \quad  \quad  \quad 
+n_1n_2^2+n_1n_3^2 +n_1n_4^2 +n_2n_3^2 +n_2n_4^2 +n_3n_4^2
\nonumber\\
\label{fourth}
&&\hspace{-.8in} \quad  \quad  \quad 
+n_1n_2n_3 +n_1n_2n_4 +n_1n_3n_4 +n_2n_3n_4 \,\,\,
-n_1 -n_2 -n_3 -n_4.
\end{eqnarray}
This recursive solution may be extended to arbitrary order and for the
general case  $\,c^{(i)}_{n+1}$ will have factors in the denominator of
\begin{eqnarray}
\label{cnp11}
&&\hspace{-.1in}
c^{(1)}_{n+1}: \quad \quad \quad   n_1 +n_2 -n_3 -n_4 \, \pm n,
\\
\label{cnp12}
&&\hspace{-.1in}
c^{(2)}_{n+1}: \quad \quad \quad   n_1 -n_2 -n_3 -n_4 \, \pm n,
\\
\label{cnp13}
&&\hspace{-.1in}
c^{(3)}_{n+1}: \quad \quad \quad   n_1 -n_2 +n_3 -n_4 \, \pm n,
\\
\label{cnp14}
&&\hspace{-.1in}
c^{(4)}_{n+1}: \quad \quad \quad   n_1 +n_2 +n_3 +n_4 \, \pm n.
\end{eqnarray}
We thus conclude that as long as there are no vanishing factors in the 
denominator  there are four distinct solutions of the equation
(\ref{okacanonical}) which are analytic at $\, t=\, 0$ and
have no arbitrary constants. These form a four dimensional
representation of the symmetry group of the equation. The statement
that the solution is analytic at $\, t= \, 0$ is the boundary condition.
This   corresponds to
the case (18) on page 7 of~\cite{guzzetti} where the subgroup of the 
monodromy group generated by
$\, M_0 M_t$ and $\, M_1$ is reducible.  

\vskip .1cm

\subsection{Arbitrary constants}
\label{arbitrary}

This recursive solution will break down at an order $n$ where
$\, c^{(i)}_n$ has a zero in the denominator. This will give a solution
only if there is a corresponding zero in the numerator which indicates
that the corresponding recursive equation is automatically satisfied
independently of the value of $\, c^{(i)}_n$ which now becomes an arbitrary
parameter that must be specified as an additional boundary condition.

There are two ways in which these vanishing factors in the numerator
can happen. Either $\, c_2^{(i)} =\, 0\,  $ or $\,  N^{(i)}_n =\, 0$. In
this paper we will apply this analysis to the correlations
$\, C(M,N)$ with $\, \nu=\, -k$. 
The behaviors of solutions of
cases 1 and 4 are quite different from cases 2 and 3 and we  treat them
separately.
 
\subsubsection{Cases 1 and 4 for $ \, T< \, T_c$. \\}
\label{cases1}

For $\, T<\, T_c\,$ the Okamoto parameters for $\, C(M, N)$  
are given by  (\ref{okam}) so that  
\begin{equation}
\hspace{-.7in}
n_1 +n_2\, = \, \, N, \quad \quad \, \, \,  {\rm and}
   \quad \quad \quad n_3 +n_4\, =\, \, 0, 
\end{equation}
and thus from (\ref{a01}) and (\ref{a04})
\begin{equation}
\label{c014m}
\hspace{-.6in} \quad 
c_0^{(1)} \, \, =\,\, \,  c_0^{(4)} \, = \,\,\,
-\frac{1}{8} \cdot \, \Bigl(N^2 -M^2 \, \, -{{1 \, +(-1)^{M+N}} \over {2}}\Bigr), 
\end{equation}
and from (\ref{a11}) and (\ref{a14}):
\begin{equation}
\label{c114m}
\hspace{-1.2in}
c_1^{(1)}\, \, =\, \, \, c_1^{(4)}\, = \, \, -\frac{M^2}{4}.
\end{equation}
Furthermore we see, from (\ref{a21}),  (\ref{a24}), (\ref{a31}) 
and (\ref{a34}), that, because of the factor of $\, n_3 +n_4$, 
the recursive solutions for cases 1 and 4, 
$\, c^{(1)}_k$ and $\, c^{(4)}_k$, will always vanish  unless there is also a
vanishing factor in the denominator. When for some $\, k$ the denominator 
in $\, c^{(1,4)}_k$ does vanish then  $\, c^{(1,4)}_k$ for 
that $\, k$ is not determined from the
recursive procedure and is an arbitrary constant. 
For $\, k=\, 2\, $ and $\, T < \, T_c\, $ we explicitly see from
(\ref{a21}) and (\ref{a24}), and for $\, k= \, 3$ from (\ref{a31}) and 
(\ref{a34}), that the factor in the denominator
\begin{equation}
\hspace{-.7in} \,\, \,\,
n_1 \, +n_2\,\,\,\, \pm \,(n_3+n_4)\,\,\,\, -k\,+1 \,\, = \,\,\,\,  N\,+1 \, \,-k, 
\end{equation}
vanishes for $\,k\, = \, \, N+1$. This pattern continues for all $\,k$ and thus 
\begin{equation}
\label{ck14mvanish}
\hspace{-.7in}
c_k^{(1)} \, = \,\, c_k^{(4)} \,= \,\,0 \quad \quad \quad  {\rm for}
    \quad   \quad  \quad   2 \, \leq \,  k  \, \leq \,  N, 
\end{equation}
and the coefficients $\, c^{(1)}_{N+1}$ and $\, c^{(4)}_{N+1}$  are arbitrary. 

To compute the coefficients $\, c^{(1,4)}_k\, $ for $\, k > \, N+1\, $
 a new recursive
solution must be computed which uses the term
$\, c_{N+1} \cdot \,  t^{N+1}\, $ as input.

Thus for $ \, t \,\rightarrow \,  \,0$
\begin{eqnarray}
&&\hspace{-.3in}\quad  \quad
h \,\,= \, \,\,
  -\frac{1}{8}\cdot \, \Bigl(N^2-M^2 \,-\frac{1 \,+(-1)^{M+N}}{2}\Bigr)
  \nonumber \\ 
  &&\hspace{-.3in} \quad  \quad  \quad  \, \,  \quad\quad  \quad
  \,-\frac{M^2}{4} \cdot  \,t \, \,\,  \, \,
+c_{N+1}\cdot \, t^{N+1} \, \,  \, \, +O(t^{N+2})
\end{eqnarray}
Thus from (\ref{shiftcmnm}) we find for $\, t \, \rightarrow\,  0$
\begin{equation}
\hspace{-.3in}
C(M,N)\, \, =\, \, \, (1-t)^{1/4} \,  \,\,
  +K(M,N)\cdot \, t^{N+1} \, \,  \, +O(t^{N+2}), 
\end{equation}
which agrees with the series expansions of $\, C(M,N)\,$ for $\, T <\, T_c$.  

\vskip .1cm

\subsubsection{Cases 2 and 3 for $ \, T > \, T_c$ with $\, M+N$ even.\\} 
\label{subsubsecCase23}

For $\, T> \, T_c$  with $\, M+N$ even we see with the Okamoto parameters
for $\, C(M,N)$ of (\ref{okap}) that
for cases 2 and 3 we have from (\ref{a02}) and (\ref{a03})
\begin{eqnarray}
\label{nc02p}
&&\hspace{-.91in}  \, \,  \,  \,  
c_0^{(2)}\,\,  = \, \, \, \frac{1}{8} \cdot \,  (M^2-N^2-1) \, \, -\frac{N}{2}, 
\quad \quad
c_0^{(3)}\,\,  =\, \,\,  \frac{1}{8} \cdot \, (M^2-N^2-1) \,\,  +\frac{N}{2}, 
\end{eqnarray}
and from (\ref{a12}) and (\ref{a13})
\begin{eqnarray}
\label{nc12p}
&&\hspace{-.5in} \quad
c_1^{(2)}\,\,  = \, \,\,   \frac{N}{4} \cdot \, \frac{N +1 -M^2}{1+N}, 
\quad \quad \quad
c_1^{(3)} \,\,  = \, \, \, \frac{N}{4}  \cdot \, \frac{N-1 +M^2}{1-N}. 
\end{eqnarray}
Furthermore the denominator in $\, c^{(2)}_{n} $ vanishes when $\, n=\, N+2\, $ 
and the denominator of $\, c^{(3)}_{n}$ vanishes when $\, n=\, N$
and thus $\, c^{(2)}_{N+2}\,$ and $\, c^{(3)}_{N}\, $ are arbitrary.
By using the definition of $ \, h$ in (\ref{hevenp}) and comparing with
the series expansions of $\, C(M,N)$ we see  for $ \, T> \, T_c$ and $\, M+N$
even that $\, C(M,N)\, $ is in case 2.

\vskip .1cm

\subsubsection{Cases 2 and 3 for $\,  {\tilde C}(M,N)\,  $ for $ \, T>\, T_c$ with
  $\, M+N$ odd \\}

For $\, T>\, T_c$ and $\, M+N$ odd we see with the Okamoto parameters
of (\ref{okapodd}) that for cases 2 and 3 we have from (\ref{a02}) and (\ref{a03})
\begin{eqnarray}
\hspace{-.8in} \quad \quad 
c_0^{(2)}\, \,  = \, \,  \, \frac{1}{8} \cdot \,  (M^2-N^2) \, -\frac{N}{2}, 
\quad \quad
c_0^{(3)}\, \,  =\, \, \,  \frac{1}{8} \cdot \, (M^2-N^2) \, +\frac{N}{2}, 
\end{eqnarray}
and from (\ref{a12}) and (\ref{a13})
\begin{eqnarray}
&&\hspace{-.6in} 
c^{(2)}_1\, \,=\, \,\, \frac{N^2-1\,-(M^2-1)N }{4 \cdot \, (1+N)}, 
\quad\quad
c^{(3)}_1\, \,=\,\, \, \frac{N^2-1\,+(M^2-1)N}{4 \cdot \, (1-N)}.
\end{eqnarray}

Furthermore the denominator in $\, c^{(2)}_{n}$ also vanishes when $n=\, N+2$
and the denominator of $\, c^{(3)}_{n}$ vanishes when $n=\, N$
and thus $\, c^{(2)}_{N+2}\,$ and $\, c^{(3)}_{N}\, $ are arbitrary.
By using the definition of $h$ in (\ref{htildep}) and comparing with
the series expansions of $ {\tilde C}(M,N)$ we see  for
$\, T>\, T_c \, $ and $\, M+N$ odd that $ {\tilde C}(M,N)$ is in case 2.

\vskip .1cm

\subsection{Determination of $\, \lambda$ when $\, \nu \, = \, -k$.}
\label{subDeterm}

It remains to determine the values of the arbitrary parameter 
which are appropriate for $\, C(M,N)$ when $\, \nu \, = \, -k$. To do
this we first  examine the behavior 
$\, C(M,N)$ at $\, t\, \rightarrow \, 0 \, \, $ for several values of
$\, M$ and $\, N$.  We consider $\, T<\, T_c$ and $\, T>\, T_c$
separately.

\vskip .1cm

\subsubsection{ $T\, <\, T_c$.\\}
\label{subsubT}
  
For $\, T< \, T_c \, $ the correlations are in case 1=4 where the arbitrary
constant is at order $\, t^{N+1}$. Several examples are
\begin{eqnarray}
\label{c01l}
&&\hspace{-.3in}
C(0,1; \, \lambda) \,\, = \, \,\,
(1-t)^{1/4} \cdot \,
\Bigl[ 1\, \, -\lambda^2 \cdot \, \left(\frac{1}{2^6}  \, t^2 \,\, +O(t^3)\right) \Bigr], 
\\
\label{c02l}
&&\hspace{-.3in}
C(0,2; \, \lambda)\, \,= \, \,\,
(1-t)^{1/4}\cdot \,
\Bigl[ 1 \, \, +\lambda^2  \cdot \, \left(\frac{1}{2^8} \, t^3 \,\,  +O(t^4)\right) \Bigr], 
\end{eqnarray}
\begin{eqnarray}
\label{c12l}
&&\hspace{-.3in}
C(1,2; \, \lambda)\,\, = \, \,\,
(1-t)^{1/4} \cdot \,
\Bigl[ 1 \,  \,-\lambda^2  \cdot \, \left(\frac{1}{2^8}\, t^3 \,\, +O(t^4)\right) \Bigr], 
\\
\label{c03l}
&&\hspace{-.3in}
C(0,3;\, \lambda)\,\, = \, \,\,
(1-t)^{1/4} \cdot \,
\Bigl[ 1 \, \, -\lambda^2  \cdot \, \left(\frac{9}{2^{14}} \, t^4 \, \, +O(t^5)\right)\Bigr], 
\\
\label{c13l}
&&\hspace{-.3in}
C(1,3; \, \lambda) \, \, = \, \, \,
(1 \,-t)^{1/4} \cdot \,
\Bigl[ 1\, \, +\lambda^2 \, \left(\frac{3\cdot   5}{2^{14}}\, t^4
   \, \, +O(t^5)\right) \Bigr], 
\end{eqnarray}
where $\,\lambda$ is chosen so that $\, \lambda=\, 1$ agrees with $\, C(M,N)$.

In general one has:
\begin{eqnarray}
\label{cmnl}
&&\hspace{-.77in} \quad 
C(M,N;\, \, \lambda) 
\nonumber \\
&&\hspace{-.67in} \quad 
\,  \,  =  \, \, \,(1-t)^{1/4} \cdot \,
\Bigl[1\,\,   +(-1)^{M+N} \cdot \, \lambda^2 \cdot \,
  \left(K_{M,N}\, t^{N+1} \,\,  +O(t^{N+2} \right))\Bigr].
\end{eqnarray}
We note for $\, \lambda\, =\, 0 \, $ that 
\begin{equation}
\hspace{-.8in}
C(M,N;\, \, 0) \, \, = \, \,  \, (1\, -t)^{1/4}, 
\end{equation}
is an exact solution to the non-linear differential equation.

\vskip .1cm

\subsubsection{$T > \, T_c\, $ for $\, M+N$ even.\\}
\label{subTMN}

These correlations are in the class 2 where the arbitrary constant is
at order $\, t^{3N/2+2}$. Several examples are
\begin{eqnarray}
&&\hspace{-1in}
  C(0,2;\, \lambda)\,\,  =\, \,\,
  (1\, -t)^{1/4} \cdot \, \frac{t}{8} \cdot  \,
  \Bigl[1 \,  \, +\frac{3}{4} \, t \, \,  +\frac{3\cdot5^2}{2^7} \, t^2
    \, \,  +\frac{5 \cdot 7^2}{2^9} \, t^3 \, \, 
+{\tilde \lambda}^2 \, \frac{3309}{2^{13}} \, t^4 \,  \,  \, +O(t^5)\Bigr]
\nonumber\\
\label{c02pl}
&&\hspace{-.5in}
= \, \,\,  (1\, -t)^{1/4} \cdot  \, \frac{t}{8} \cdot \,
\Bigl[ {}_2F_1\Bigl([\frac{3}{2}, \frac{3}{2}], \, [3], \, \, t\Bigr)
\, \, + \lambda^2 \cdot \, \left(\frac{3}{2^{14}} \, t^4 \, \, +O(t^5)\right)\Bigr], 
\end{eqnarray}
\begin{eqnarray}
&&\hspace{-1in}
C(1,3; \, \lambda) \,\,  = \, \,\,
(1\, -t)^{1/4} \cdot \, t^{3/2} \cdot \, \frac{1}{16}
\cdot  \,
\Bigl[1 \,  \,+\frac{3\cdot 5}{2^4} \, t \, \, +\frac{3\cdot 5\cdot 7}{2^7} \, t^2
\,  \,+\frac{3\cdot 5\cdot 7^2}{2^{10}} \, t^3 \,
\nonumber\\
&&\hspace{-.5in} \quad  \quad \quad  \quad 
 +\frac{3^3 \cdot 5 \cdot 7 \cdot 11}{2^{14}} \,\, t^4 \,  \,  \, 
+{\tilde\lambda}^2 \,  \frac{297315}{2^{19}} \,\, t^5 \, \, \,  \, +O(t^6)\Bigr]
\\
\label{c13pl}
&&\hspace{-.5in}
= \, \, \,\, 
(1\, -t)^{1/4} \cdot \, t^{3/2} \cdot \, 
\frac{1}{2^4} \cdot \,
\Bigl[ {}_2F_1\Bigl([\frac{3}{2},\frac{5}{2}], \,[4],  \, \, t\Bigr) \,\,
+\lambda^2 \, \left(\frac{3}{2^{18}} \,\, t^5 \,\, \,+O(t^6)\right)\Bigr]
\nonumber
\end{eqnarray}
\begin{eqnarray}
&&\hspace{-1in} \quad  \quad 
C(0,4;\, \lambda)\,\, = \, \,\,
(1\, -t)^{1/4} \cdot \, t^2  \cdot\, \frac{3}{2^7}
 \cdot \, \Bigl[1\,\,  +\frac{5}{2^2} \,\, t \, \,
 +\frac{5 \cdot 7^2}{3 \cdot 2^6} \,\, t^2\,\,
  +\frac{3^2 \cdot 5 \cdot 7}{2^8} \,\,  t^3
\nonumber\\
&&\hspace{-.5in} \quad \quad 
+\frac{3^2\cdot 5 \cdot 7 \cdot 11^2}{2^{15}}\,\,  t^4 \, \,  \, 
+\frac{7 \cdot 11^2 \cdot 13^2}{2^{17}}\, \,  t^5 \, \, 
+{\tilde\lambda}^2 \cdot  \, \frac{5 \cdot 429431}{2^{21}} \, \,  t^6 \Bigr]
\\
\label{c04pl}
&&\hspace{-.5in} \quad 
= \, \, \, (1\, -t)^{1/4}  \cdot \, t^2  \cdot \, \frac{3}{2^7} \cdot \, 
\Bigl[ {}_2F_1\Bigl([\frac{5}{2},\frac{5}{2}], \, [5],  \, \, t\Bigr)
  \,  \,+\lambda^2 \cdot \, \left(\frac{5}{2^{20}} \, \, t^6
   \,\, \, +O(t^7)\right)\Bigr],
\nonumber
\end{eqnarray}
where both $\, \lambda$ and $\, {\tilde \lambda}$ are chosen such that, when
$\, \lambda=\, 1$ and ${\tilde \lambda}=\, 1$, there is agreement with $\, C(M,N)$.
In general:
\begin{eqnarray}
&&\hspace{-1in} \quad  \quad  \quad 
  C(M,N;\, \, \lambda) \, \, \, = \, \,\,\,
  \nonumber \\
&&\hspace{-1in} \quad \quad  \quad \quad  \quad 
%   \, = \, \,\,\,
(1\, -t)^{1/4} \cdot  \, t^{N/2} \cdot  \, K^{(1)}_{M,N}
\cdot \,
\Bigl[{}_2F_1\Bigl([\frac{N-M+1}{2},\frac{N+M+1}{2}], \, [N+1], \, \, t\Bigr)
\nonumber\\
\label{cmnpl}
&&\hspace{-.5in} \quad  \quad \quad  \quad 
+\lambda^2 \cdot \, \Bigl(K^{(2)}_{M.N} \cdot \, t^{N+2} \, \, \, +O(t^{N+3})\Bigr)\Bigr].
\end{eqnarray}

We note that $\, {\tilde \lambda} \, $ and $\, \lambda \, $ are not the same. When
$\, {\tilde \lambda}\, =\,\,  0$ the term $\, O(t^{N+3})$ does not in general
vanish. This is in contrast with the case $\, \lambda\, =\,\,  0$ where, for some constant $\, \rho$
\begin{eqnarray}
&&\hspace{-.8in} \quad 
C(M,N;\,\, 0)  \,\, \, = \, \,\,
\nonumber \\
&&\hspace{-1in} \quad \quad  \quad  \quad 
\rho \cdot \, (1\, -t)^{1/4} \cdot \, t^{N/2} \cdot \,
  {}_2F_1\Bigl([\frac{N-M+1}{2},\frac{N+M+1}{2}], \, [N+1], \, \, t\Bigr), 
\end{eqnarray}
is an exact solution to the nonlinear differential  equation. The constant 
$\, K^{(1)}_{M,N} \, $ is a normalization constant
which cannot be determined from the non linear equation.

In the specific examples (\ref{c01l})-(\ref{c13l}),
(\ref{c02pl})-(\ref{c04pl}) the numerical coefficient of $\, \lambda^2\,$ has
been chosen so that $\, \lambda  \, = \, \,  1\,$ is the desired correlation where
$\, C(M,N)$ is given as a finite homogenerous polynomial in $\, {\tilde K}$
and $\,{\tilde E}$, and in the case of $\, C(0,N)$ as an $\, N \, \times\,  N$
Toeplitz determinant. However, in general no  explicit formula for $\, K_{M,N}$ 
or $\, K^{(2)}_{M,N}$ is known which allows $\, \lambda \, =\,  \, 1$
to reduce to the desired result $\, C(M,N)$.

\vskip .1cm

\subsubsection{$T > \, T_c\, \, $ for $ \, M+N$ odd.\\}
\label{subTMN2}

Similarly, one can see that
\begin{eqnarray}
&&\hspace{-1in} \quad \quad
\tilde{C}(M,N;\, \, \lambda)
\nonumber \\
&&\hspace{-1in} \quad \quad \quad \quad
\, = \, \,\,\, 
(1\, -t)^{-1/4} \cdot \, t^{N/2} \cdot \, K^{(1)}_{M,N}
\cdot \,
\Bigl[{}_2F_1\Bigl([\frac{N-M}{2},\frac{N+M}{2}], \, [N+1], \, \, t\Bigr)
\nonumber\\
\label{cmnpl}
&&\hspace{-.5in} \quad \quad \quad \quad \quad 
+\lambda^2 \cdot \, \Bigl(K^{(2)}_{M.N} \cdot \, t^{N+2} \, \, \, +O(t^{N+3})\Bigr)\Bigr],
\end{eqnarray}
and, for some constant $\, \rho$, that
\begin{equation}
\hspace{.01in} \, \, 
\tilde{C}(M,N;\,\, 0)\,\, = \, \,\,
\rho \cdot \, (1\, -t)^{-1/4} \cdot \, t^{N/2} \cdot \,
{}_2F_1\Bigl([\frac{N-M}{2},\frac{N+M}{2}], \, [N+1], \, \, t\Bigr),
\end{equation}
is an exact solution to the nonlinear differential equation (\ref{eqnp}).

\vskip .4cm
\vskip .4cm
\vskip .4cm

\vspace{.1in}
\pagebreak

\end{document}